\documentclass[]{aastex631}

\begin{document}

\title[Polarization of Quasar 1604+159]{Evolution of magnetic field of the Quasar 1604+159 at pc scale}

\author[0000-0002-5398-1303]{Xu-Zhi Hu}
\thanks{E-mail: hxz@shao.ac.cn}
\affiliation{Shanghai Astronomical Observatory, Chinese Academy of Sciences, Shanghai 200030, People's Republic of China}
\affiliation{School of Physical Science and Technology, ShanghaiTech University, Shanghai 201210, People's Republic of China}
\affiliation{School of Astronomy and Space Science, University of Chinese Academy of Sciences, Beijing 100049, People's Republic of China}

\author[0000-0002-1992-5260]{Xiaoyu Hong}
\affiliation{Shanghai Astronomical Observatory, Chinese Academy of Sciences, Shanghai 200030, People's Republic of China}
\affiliation{School of Physical Science and Technology, ShanghaiTech University, Shanghai 201210, People's Republic of China}
\affiliation{School of Astronomy and Space Science, University of Chinese Academy of Sciences, Beijing 100049, People's Republic of China}
\affiliation{Key Laboratory of Radio Astronomy, Chinese Academy of Sciences, 210008 Nanjing, People's Republic of China}

\author[0000-0002-1992-5260]{Wei Zhao}
\affiliation{Shanghai Astronomical Observatory, Chinese Academy of Sciences, Shanghai 200030, People's Republic of China}
\affiliation{Key Laboratory of Radio Astronomy, Chinese Academy of Sciences, 210008 Nanjing, People's Republic of China}

\author[0000-0002-1908-0536]{Liang Chen}
\affiliation{Shanghai Astronomical Observatory, Chinese Academy of Sciences, Shanghai 200030, People's Republic of China}
\affiliation{Key Laboratory for Research in Galaxies and Cosmology, Shanghai Astronomical Observatory, Chinese Academy of Sciences, Shanghai 200030,\\
People’s Republic of China}

\author[0000-0001-9036-8543]{Wei-Yang Wang}
\affiliation{School of Astronomy and Space Science, University of Chinese Academy
of Sciences, Beijing 100049, People's Republic of China}
\affiliation{School of Physics and State Key Laboratory of Nuclear Physics and Technology, Peking University, Beijing 100871, People's Republic of China}
\affiliation{Kavli Institute for Astronomy and Astrophysics, Peking University, Beijing 100871, People's Republic of China}

\author[0000-0003-3454-6522]{Linhui Wu}
\affiliation{Shanghai Astronomical Observatory, Chinese Academy of Sciences, Shanghai 200030, People's Republic of China}



\begin{abstract}

We have analyzed the total intensity, spectral index, linear polarization, and RM distributions at pc scale for the quasar 1604+159.
The source was observed at 5.0, 8.4, and 15.4 GHz in 2002, and 4.6, 5.1, 6.0, 7.8, 12.2, 15.2, and 43.9 GHz in 2020 with the American Very Long Baseline Array (VLBA).
Combining the polarization results of Monitoring Of Jets in Active galactic nuclei with VLBA Experiments (MOJAVE) at 15 GHz from 2009 to 2013, we studied the evolution of magnetic field at pc scale for the source.
We detected a core-jet structure.
The jet extends to the distance of $\sim 25$ mas from the core at a direction of $\sim 66^{\circ}$ north by east.
The shape of the jet derived from 15 GHz data varies slightly with time and could be described by a straight line.
Based on the linear polarization distribution in 2002, we divided the source structure into the central region and the jet region.
In the jet region, we find the polarized emission varies with time.
The flatter spectral index values and EVPA direction indicate the possible existence of shocks, contributing to the variation of polarization in the jet with time.
In the central region, the derived core shift index $k_{r}$ values indicate that the core in 2002 is close to the equipartition case while deviating from it in 2020.
The measured magnetic field strength in 2020 is two orders of magnitude lower than that in 2002.
We detected transverse RM gradients, evidence of a helical magnetic field, in the core.
The polarized emission orientates in general toward the jet direction in the core.
At 15 GHz, in the place close to the jet base, the polarization direction changes significantly with time from perpendicular to parallel to the jet direction.
The evolution of RM and magnetic field structure are potential reasons for the observed polarization change.
The core $\rm |RM|$ in 2020 increases with frequency following a power law with index $a = 2.7 \pm 0.5 $, suggesting a fast electron density fall-off in the medium with distance from the jet base.
\end{abstract}



\section{Introduction}
Synchrotron emission is believed to be the main radio emission mechanism of jets of active galactic nuclei (AGN).
It results from the acceleration of the ionized electrons in the magnetic ($\textit{\textbf{B}}$) field and is highly polarized with theoretical maximum fractional linear polarization of 75 percent \citep{1970ranp.book.....P}.

The observed polarized emission strongly connects with the adjacent $\textit{\textbf{B}}$ field, which is one of the key factors contributing to the propagation of jets.
$\textit{\textbf{B}}$ field wound up by rotating accretion disk or central supermassive black hole (SMBH) extends with jets to distance of parsec (pc) scale even up to kilo-parsec (kpc) scale \citep{2015A&A...583A..96G}.
Thus, the study of the $\textit{\textbf{B}}$ field at pc scale is important for the investigation of the physical origin of the propagation of jets.

$\textit{\textbf{B}}$ field could be detected by linear polarization observation \citep[e.g.][]{1999ApJ...518L..87A,2000MNRAS.319.1109G,2010ApJ...723.1150P,2014MNRAS.438L...1G}.
According to \citet{1970ranp.book.....P}, the projected components of $\textit{\textbf{B}}$ field onto the plane of the sky are orthogonal to the polarization angle $\chi$ in optically thin regions, while parallel in optically thick regions.
This parallel relation happens at an optical depth $\tau \backsimeq 6-7$ \citep{2018Galax...6....5W}.

The information on the $\textit{\textbf{B}}$ field could also be obtained through  Rotation Measure (RM).
A set of linearly polarized electromagnetic waves emitted by an AGN jet would experience a rotation of polarization direction when moving through ionized materials, called Faraday Rotation, which could be described with RM,
\begin{equation}
    \label{RM}
    \chi_{\rm obs} - \chi_{\rm o} = \frac{e^3\lambda^2}{8\pi^2\epsilon_om^2c^3} \int n_e\textbf{\textit{B}}\cdot d\textbf{\textit{l}} \equiv \rm RM\lambda^2
\end{equation}
\citep{1975clel.book.....J}, where $\chi_{\rm obs}$ is the observed polarization angle, $\chi_{\rm o}$ is the intrinsic polarization angle, $\lambda$ is the wavelength, $n_{e}$ is the density of the electrons in the medium, and $\textbf{\textit{B}}\cdot d\textbf{\textit{l}}$ is the projected $\textit{\textbf{B}}$ filed component on the line of sight (LoS) multiplied by the integral element $d\textbf{\textit{l}}$.

Thus, the projected components of the $\textit{\textbf{B}}$ field on the LoS  could be detected through RM \citep[]{2002PASJ...54L..39A,2005ApJ...626L..73Z,2012AJ....144..105H}.
Further, detection of transverse RM gradients across jets provides strong evidence of helical $\textit{\textbf{B}}$ field  \citep[e.g.][]{2004MNRAS.351L..89G,2008ApJ...682..798A,2010ApJ...720...41A,2008MNRAS.384.1003G,2009ApJ...694.1485K,2009MNRAS.400....2M,2010MNRAS.402..259C, 2012AJ....144..105H,2015MNRAS.450.2441G,2018A&A...612A..67G}.
\cite{2011MNRAS.415.2081C} concluded that a helical $\textit{\textbf{B}}$ field could also produce significant transverse asymmetries of profiles of fractional linear polarization and spectral index.

Blazars, which include Flat-Spectrum Radio Quasar and BL Lac objects
, serve as a strong population of AGN due to the condition that their jet axes orientate close to the LoS of the observer.
Emission received would experience an enhancement of power due to the Doppler boosting effect, which makes blazars better candidates to study the $\textit{\textbf{B}}$ field in jets of AGN.

Source 1604+159 (J1607+1551) has been classified as Low-spectral peaked (LSP) Quasar with redshift z = 0.4965 \citep{2008ApJS..175..297A}. 
It has a Luminosity Distance of $ D_{L}$ = 2799 Mpc and a linear scale of 6.06 pc/mas. 
It was detected by both $\gamma-ray$ detector EGRET \citep{1995ApJS..101..259T} and Fermi \citep{2010ApJS..188..405A,2012ApJS..199...31N}.
It has flat radio spectral and low radio variability \citep{2014A&A...572A..59M,2022AstBu..77..361S}.
It is a core–jet source, from VLBI up to 8.4 GHz VLA-A scales (JVAS) \citep{2006MNRAS.368.1411A}.
\citep{2019ApJ...874...43L} analyzed its jet speed at pc scale and found a maximum 71 $\pm$ 14 $\rm \mu as/y$ (2.09 $\pm$ 0.41 c) based on 3 moving features.

\citet{2023MNRAS.520.6053P} and \citet{2023MNRAS.523.3615Z} have studied its evolution of linear polarization at 15 GHz spanning from 2009 to 2013.
They have found that the jet of 1604+159 has a slight S-shape and the polarization direction or electric vector position angle (EVPA) in general orients towards the jet direction from the VLBI core to a distance of about 4 mas from the core.
This means stable toroidal components of the $\textit{\textbf{B}}$ field exist during the propagation of the jet at pc scale.
In this work, we extend the frequency spanning to 4.6 $\sim$ 43.9 GHz and the time spanning to $\sim$ 18 years with two multi-frequency observations
We further investigate the evolution of the $\textit{\textbf{B}}$ field and surrounding medium for the quasar 1604+159 through distributions of spectral index, linear polarization, and RM.

\section{Observation and Data reduction}
\subsection{Observations}
We organized two multi-frequencies dual-polarization observations for the quasar 1604+159 with the Very Long Baseline Array (VLBA), which are the project BH096 (2002-Jul-20) and the project BH228A (2020-Nov-20).

Project BH096 has 3 frequencies centered at 5.0, 8.4, and 15.4 GHz.
Each frequency has 4 intermediate frequencies (IFs) and 32 MHz bandwidth.
The data rate is 128 Mbits $\rm s^{-1}$.
The on-source time is 56 mins.

Project BH228A has 7 frequencies spanning 3 bands with 4 in the C band, 2 in the Ku band, and 1 in the Q band, with the DDC personality of the RDBE.
Each frequency in the C band has 2 IFs and 128 MHz bandwidth.
Frequencies in the Ku band Q band have 4 IFs and 256 MHz bandwidth.
The data rate is 2 Gbits $\rm s^{-1}$.
The on-source time is 92 mins.
The information on the observations is listed in Table~\ref{tab:observation}.

\label{sec:maths} % used for referring to this section from elsewhere
\subsection{Data Reduction}
The VLBA raw data was correlated using the correlator at the Array Operations Center of National Radio Astronomy Observatory (NRAO) in Socorro with an average time\textbf{ }of 2 s.
The initial calibration was conducted using the Astronomical Imaging Processing Software (AIPS) package \citep[]{2003ASSL..285..109G} with the standard procedure.
Each frequency was calibrated separately.

For BH228A, the gain values of the C band were not attached correctly.
We updated GC tables for the C band data with the appropriate gain values after loading the data into the AIPS with the task ‘FITLD' following the suggestion provided by NRAO \footnote{\href{https://science.nrao.edu/facilities/vlba/data-processing/vlba-7ghz-flux-density-scale}{https://science.nrao.edu/facilities/vlba/data-processing/vlba-7ghz-flux-density-scale}}.

After inspection of the data, Fort Davis (FD) and Los Alamos (LA) were chosen as the reference antenna for BH096 and BH228A, respectively.
The a-prior amplitude calibrations were done using the information for all antennas in the gain curve (GC), system temperature (TY), and weather (WX) tables to correct the atmospheric opacity.
The ionospheric delay was corrected via total electron content measurements from global positioning system monitoring.
The phase contributions from the antenna parallactic angles were removed before any other phase corrections were applied.
The delay search, bandpass calibration, and RL delay were performed with the source 3C 279 for BH096, and J1751+0939 for BH228A.

The instrumental polarization (D-terms) solutions were solved using the task `LPCAL’. 
Source J1407+2827 (OQ208) served as D-terms calibrator for BH096.
In BH228A, the compact polarized source J1751+0939 with good parallactic angle coverage during the observation was used.

During the calibration of the EVPA of BH096, the compact polarized source J2253+1608 was used for 5.0 and 8.4 GHz, and 3C 279 for 15.4 GHz.
The reference EVPAs were obtained from the Very Large Array (VLA) polarization monitoring program \footnote{\href{http://www.vla.nrao.edu/astro/calib/polar/}{http://www.vla.nrao.edu/astro/calib/polar/}} and Monitoring Of Jets in Active galactic nuclei with VLBA Experiments (MOJAVE) 2cm Survey \citep{2018ApJS..234...12L}.
The nearest epoch for the frequencies 5.0 GHz and 8.5 GHz was 2002 Jul 10. 
Differences between the reference and our integrated EVPAs were $29.8^{\circ}$ and $-41.9^{\circ}$ with errors $2.6^{\circ}$ and $2.0^{\circ}$ for 5.0 GHz and 8.4 GHz, respectively.
The nearest epoch for the frequency 15.4 GHz was 2002-July-19.
The difference between the reference and our integrated EVPAs was $-21.9^{\circ}$ with error $3.0^{\circ}$ \citep{2012AJ....144..105H}.

\begin{figure}
	\includegraphics[width=0.5\linewidth]{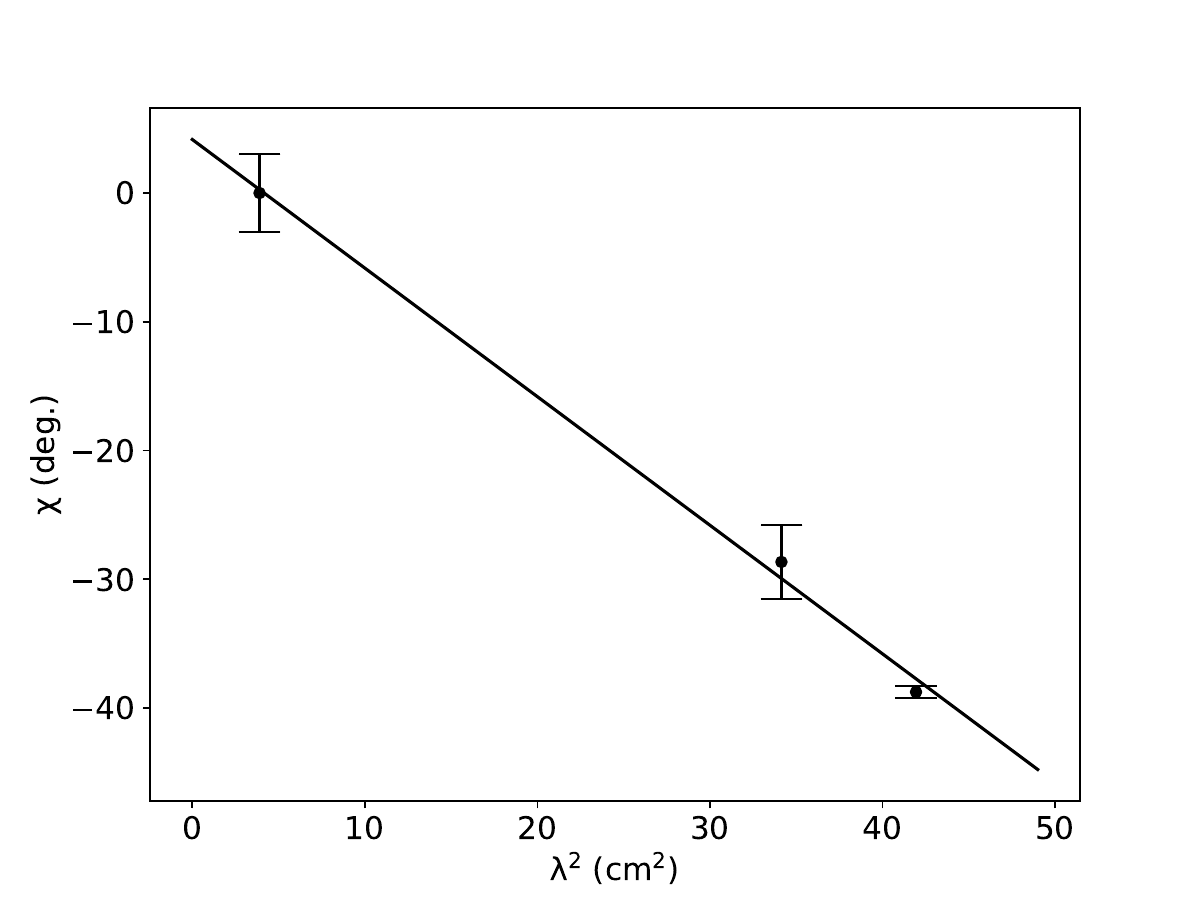}
    \centering
    \caption{EVPA calibration for the project BH228A. The plot shows EVPA - $\lambda^{2}$ law for the EVPA calibrator 2200+420 at 4.6, 5.1, and 15.4 GHz.
    The integrated RM is $\rm -174 \pm 10$ $\rm rad m^{-2}$
    The correct values of EVPA at 6.0, 7.8, and 12.2 GHz were obtained by interpolation in this law.}
    \label{fig:2200 RM}
\end{figure}

We arranged two EVPA calibrators for BH228A, source 3C 286 and 2200+420.
The source 3C 286 was used for the data of 4.6 and 5.1 GHz \citep{1997A&A...325..479C}.
The data of 15.2 GHz was calibrated using the MOJAVE 2cm Survey results of the source 2200+420 and the nearest epoch was 2020-Nov-29.
After the EVPA calibration of 4.6 and 5.1 GHz, we fitted a linear model to EVPA - $\lambda^{2}$  for 2200+420 (as shown in Fig.~\ref{fig:2200 RM}) and interpolated for other frequencies except for 43.9 GHz data to obtain the correct EVPA.
This interpolation method of EVPA calibration was also used in other studies including \citet{2009MNRAS.393..429O} and \citet{2022MNRAS.510.1480K}.
The 43.9 GHz data was calibrated with the results of VLBA Boston U \footnote{\href{https://www.bu.edu/blazars/BEAM-ME.html}{https://www.bu.edu/blazars/BEAM-ME.html}}.
The calibration errors are $\sim 3^{\circ}$.
Detailed information on calibrations is listed\textbf{ }in Table~\ref{tab:observation}.

Imaging and self-calibration were carried out in the Difmap package \citep{1997ASPC..125...77S}.
The \textit{I}, \textit{Q}, and \textit{U} distributions were then obtained using the fully self-calibrated visibilities with natural weighting.
For the C band data of BH228A, according to the same suggestion above, as the source 1604+159 is strong enough for self-calibration, we first created a clean model of the source excluding the antennas Pie Town (PT) and North Liberty (NL) in Difmap.
Then we computed the self-calibration solutions based on that model with the task `CALIB' in AIPS, and applied the solutions to all antennas with the task `CLCAL'.

The distributions of the polarization intensity ($\ p = \sqrt{Q^2+U^2}$), polarization fraction (i.e. $m=p/I$), and EVPAs were calculated pixel by pixel. 
The corresponding uncertainties were obtained using the approach of \citet{2012AJ....144..105H}, which takes into account the effect of the D-terms for each pixel. 


\begin{figure*}
	\includegraphics[width=\linewidth]{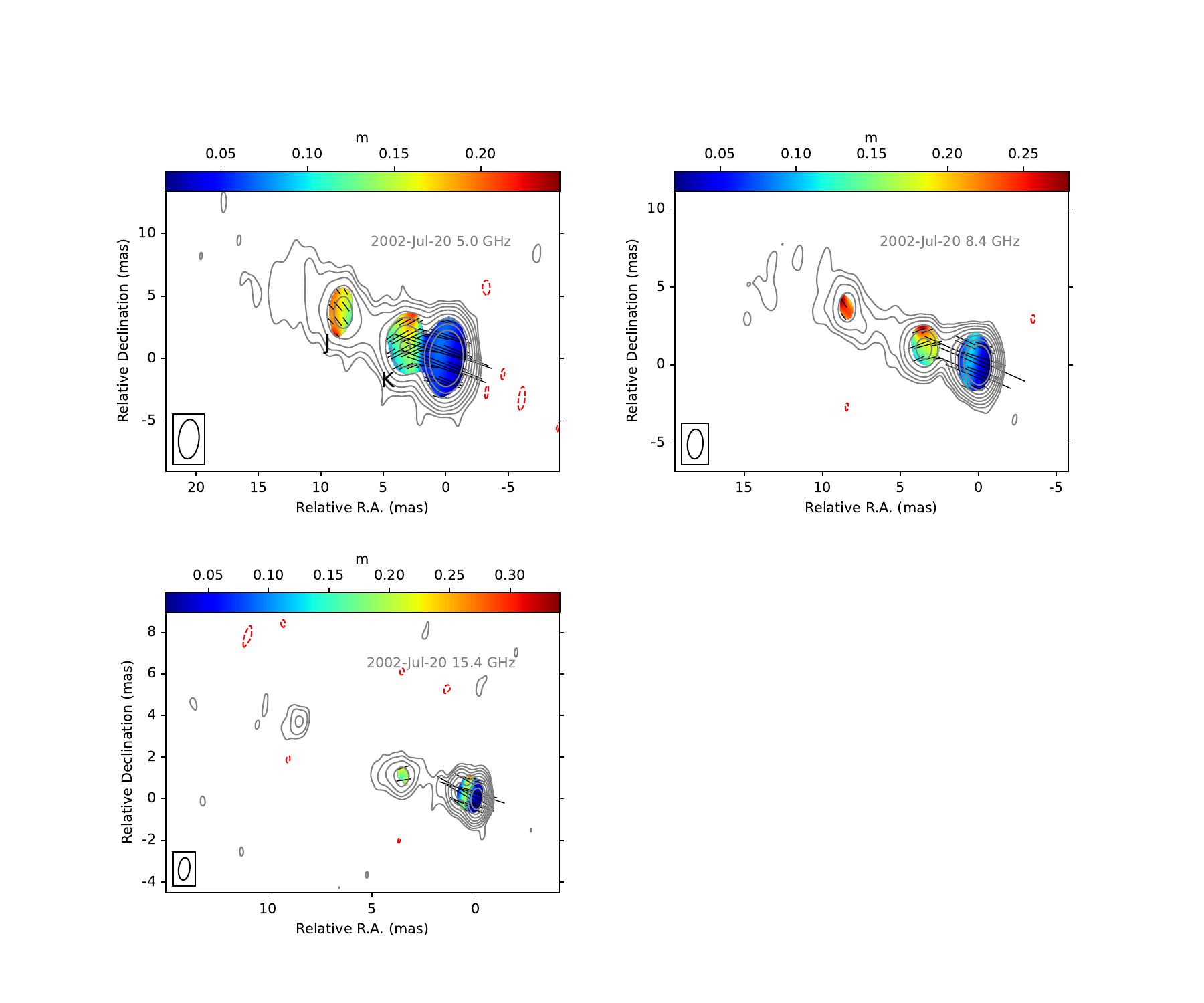}
    \caption{Distributions of linear polarization (sticks) and fractional polarization (color) for the 5.0, 8.4, and 15.4 GHz observation on 2002-Jul-20 superimposed on the corresponding total intensity contours.
    Contours are $3\sigma \times (-1, 1, 2 ,4, 8, 16, 32, 64, 128)$.}
    \label{fig:BH096 pol}
\end{figure*}

\begin{figure*}
	\includegraphics[width=\linewidth]{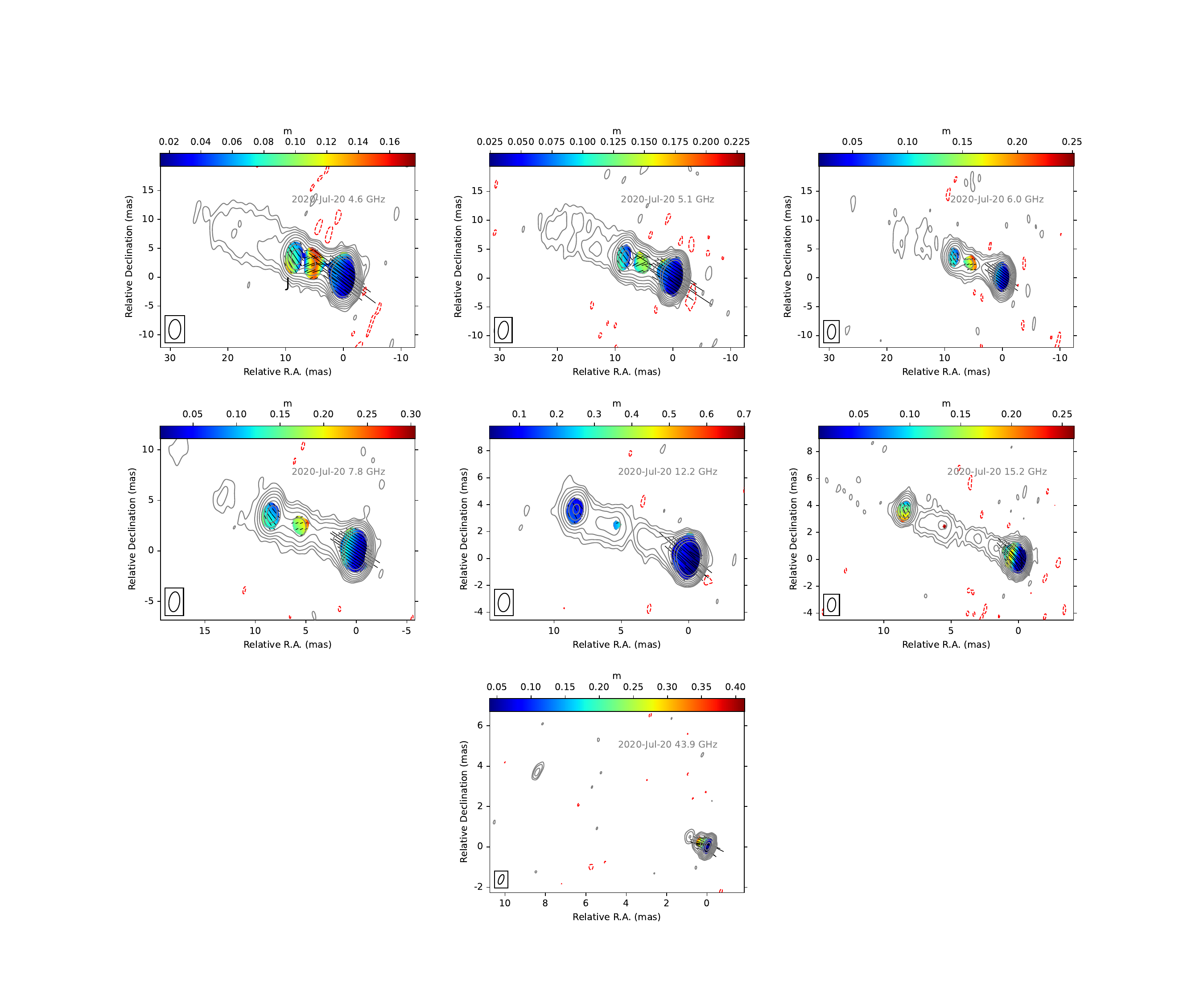}
    \caption{Distributions of linear polarization (sticks) and fractional polarization (color) for the observation on 2020-Nov-20 superimposed on the corresponding total intensity contours.
    Contours are $3\sigma \times (-1, 1, 2 ,4, 8, 16, 32, 64, 128)$.}
    \label{fig:BH228A pol}
\end{figure*}

\begin{figure}
	\includegraphics[width=0.5\linewidth]{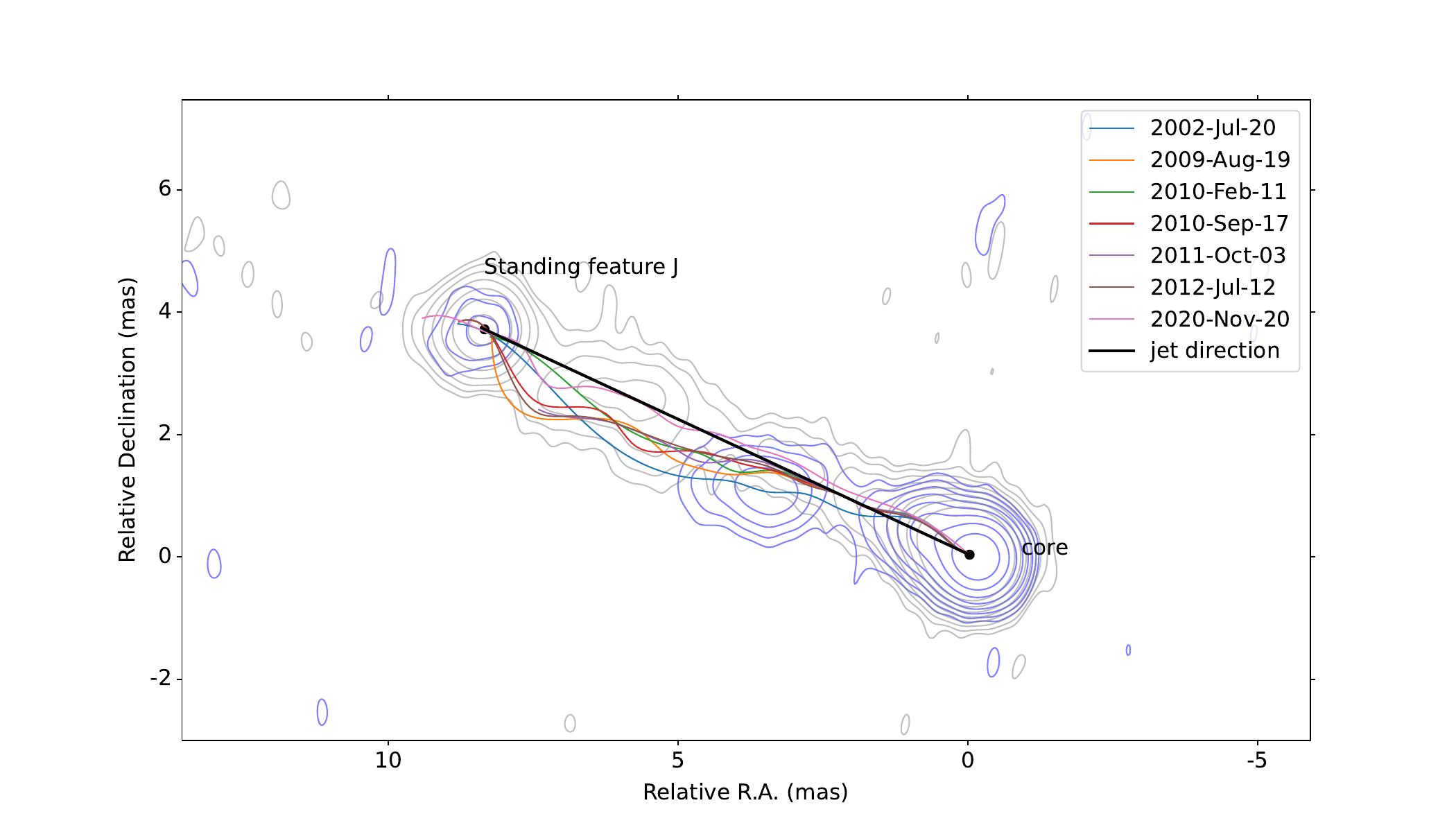}
    \centering
    \caption{
    Streamline derived to represent the jet shape for 15 GHz data from 2002 to 2020 with 2002 in blue and 2020 in pink.
    Contours are the 15 GHz total intensity (2002 in blue, 2020 in grey) with $3\sigma \times (1, 2 ,4, 8, 16, 32, 64, 128)$.
    The jet propagates along the direction between the core and the standing feature J.
    There is a hint for the jet from the core to the $\sim$ 6 mas place that the jet rotates clockwise with time.
    }
    \label{fig:Ridge Line}
\end{figure}

\begin{figure*}
	\includegraphics[width=\linewidth]{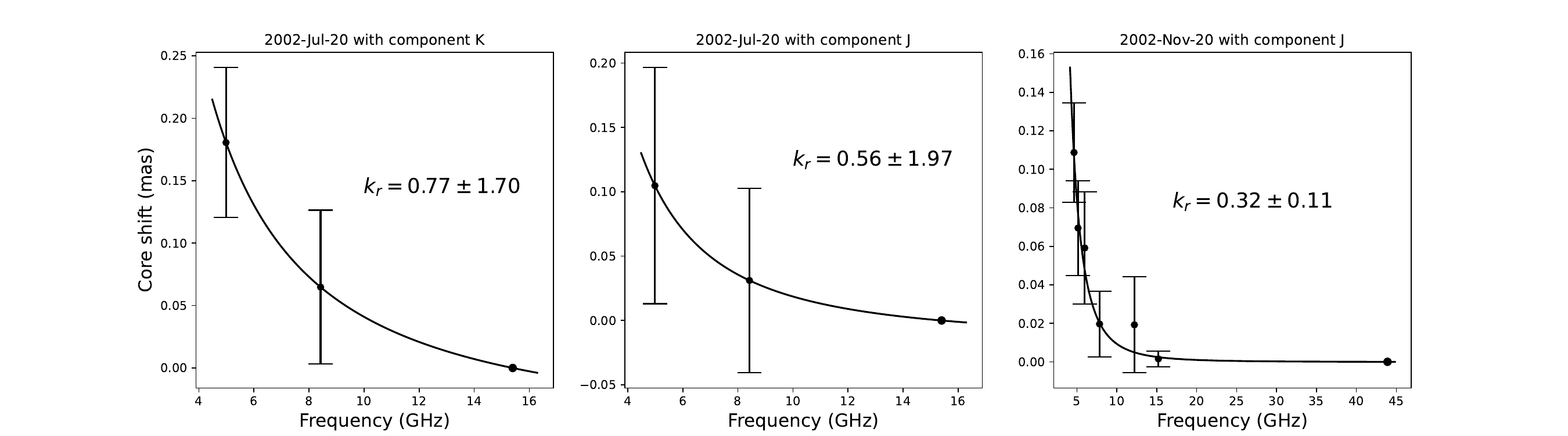}
    \caption{Core shift as a function of frequency with 15.4 GHz and 43.9 GHz as a reference frequency for observations on 2002-Jul-20 and 2020-Nov-20, respectively.
    In 2002, compact components K and J were used.}
    \label{fig:1604 coreshift}
\end{figure*}

%Shape evolution of the jet.
%The red solid line is fitted for the 15.4 GHz observation on 2002-Jul-20, the blue line is for 15.2 GHz the observation on 2020-Nov-20.
%Contours are the 4.6 GHz total intensity with $3\sigma \times (-1, 1, 2 ,4, 8, 16, 32, 64, 128)$.
%The green solid line divides the source structrue into the central region and the jet region based on the polarization distribution of 5.0 GHz in 2002.
%As the jet continues to propagate as a line after through the standing feature J, we took a straight line (black solid line) as its ridge line when sampling for spectral index and EVPA with the start point `S'. 

\begin{figure}
	\includegraphics[width=\linewidth]{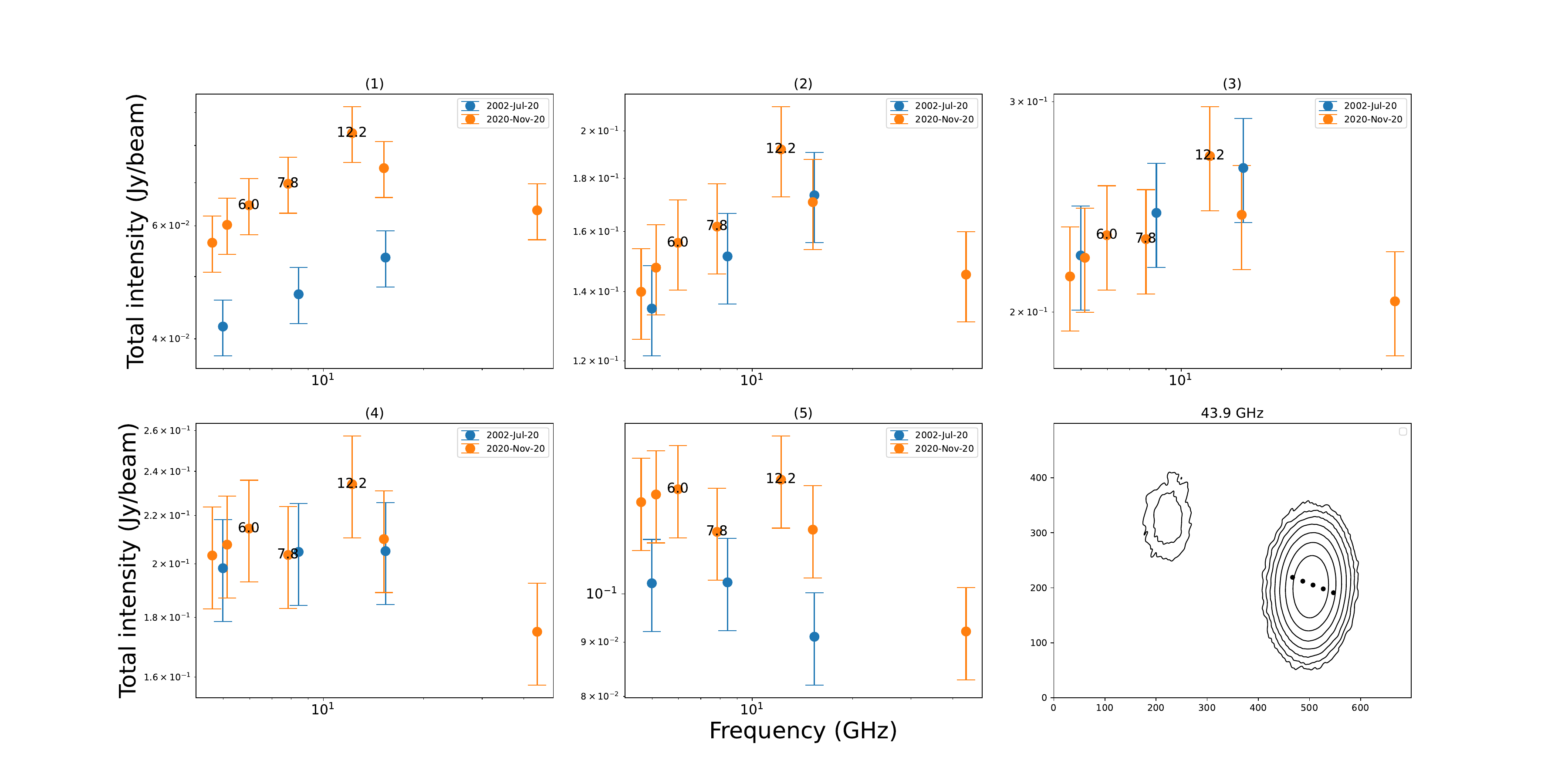}
    \centering
    \caption{A plot of total intensity against frequency for 5 points (roughly the same interval $\sim$ 0.6 mas) sampled along the jet direction from the central region (shown in the last map) for the multi-frequency observations in 2002 (blue) and 2020 (orange).
    The titles (1)-(5) indicate the positions with larger numbers corresponding to the distance further from the jet base.
    The spectra of the sample show complex change, indicating a potentially complex physical environment in the central region for the source.}
    \label{fig:1604 I vs frequency}
\end{figure}

\begin{figure*}
	\includegraphics[width=\linewidth]{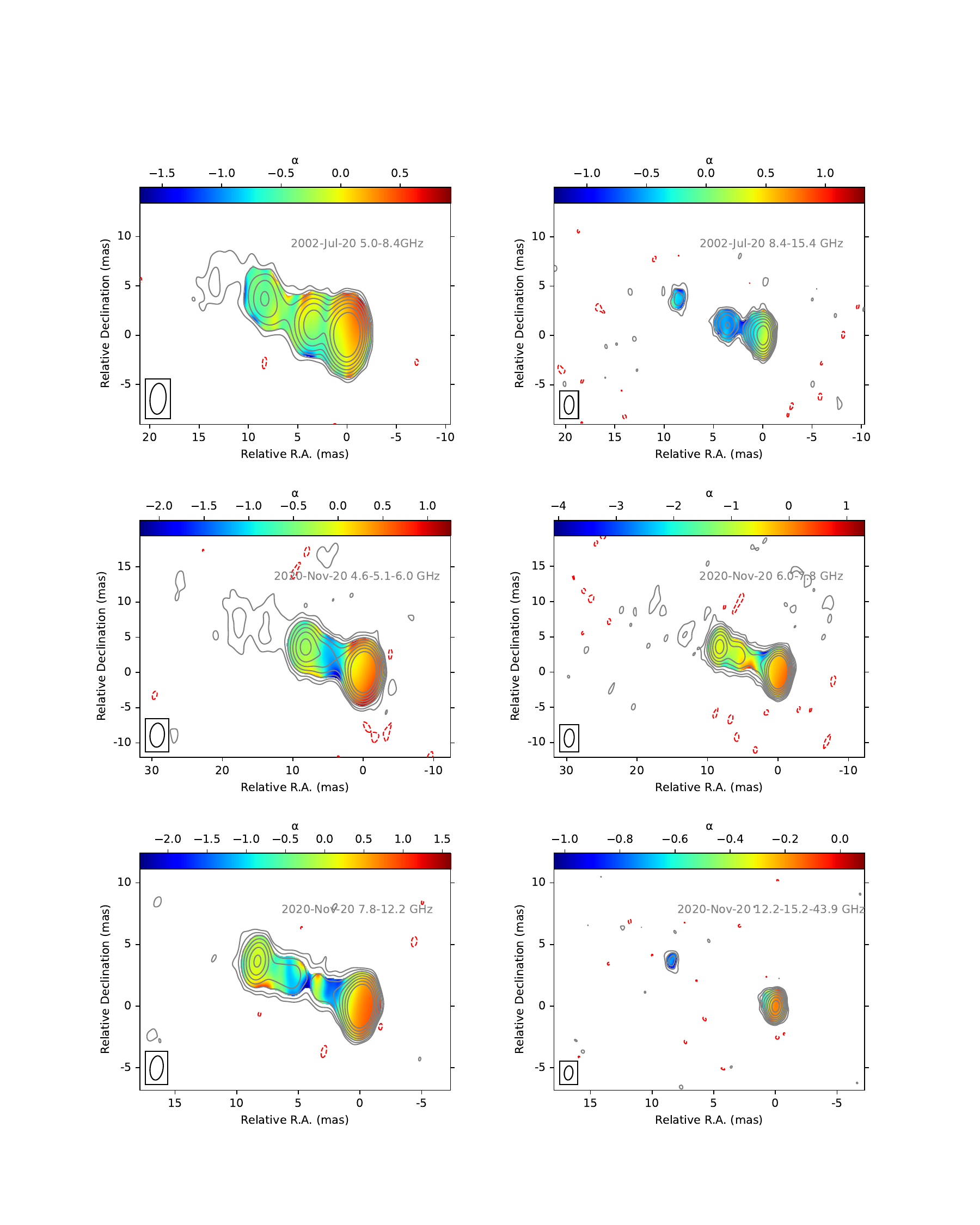}
    \caption{Distributions of the spectral index for observations on 2002-Jul-20 and 2020-Nov-20.
    Color is the spectral index ($\alpha$).
    Contours are the highest frequency total intensity at given frequency intervals with $3\sigma \times (-1, 1, 2 ,4, 8, 16, 32, 64, 128)$.}
    \label{fig:1604 Spectral index}
 \end{figure*}

\begin{figure*}
	\includegraphics[width=\linewidth]{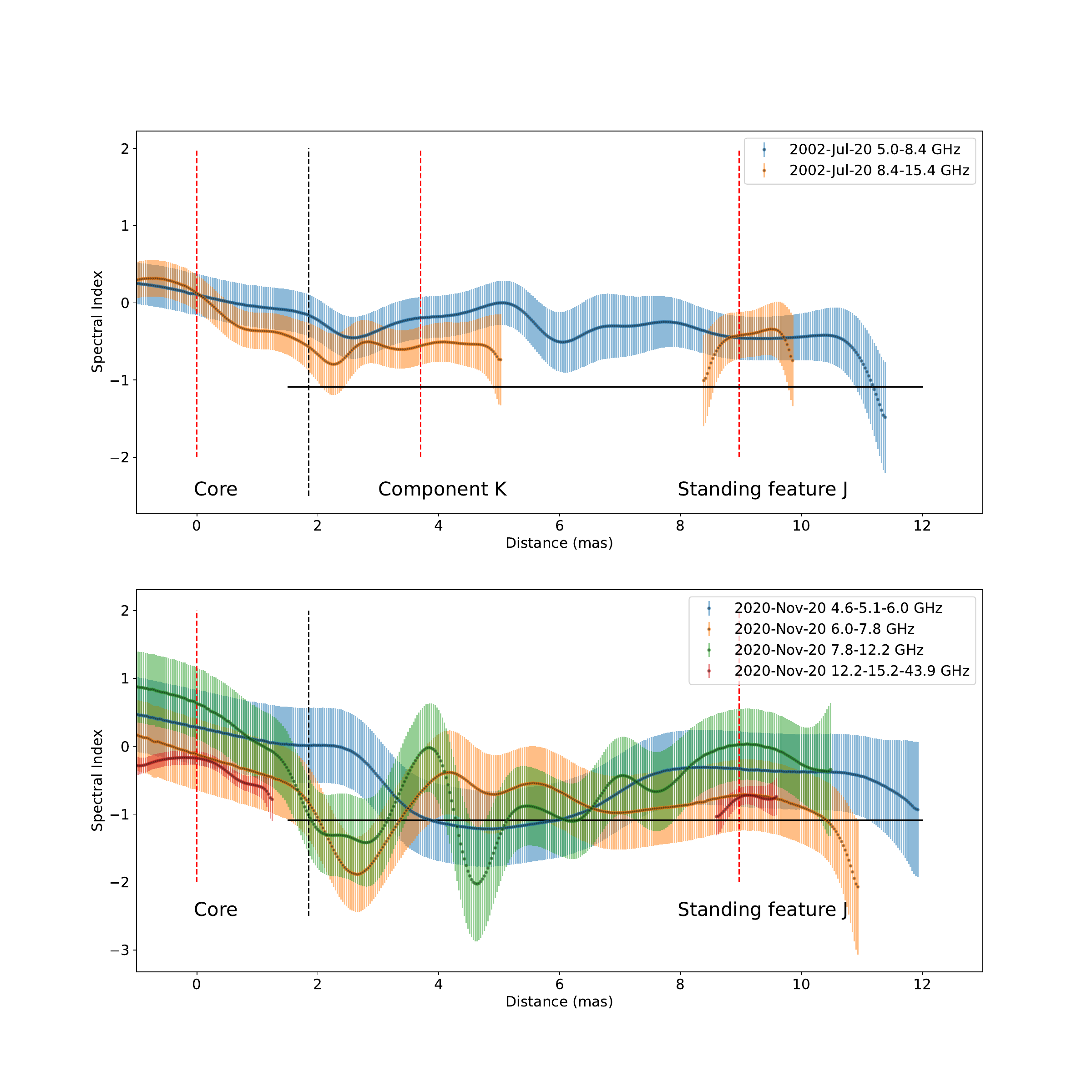}
    \caption{Distributions of the spectral index along the jet ridge line for observations on 2002-Jul-20 and 2020-Nov-20.
    The black dashed line divides the source into the central region and the jet region.
    The black solid line indicates the mean spectral index value from the jet ridge line for quasars \citep{2014AJ....147..143H}.}
    \label{fig:SPIX along the jet}
\end{figure*}

\begin{figure*}
	\includegraphics[width=\linewidth]{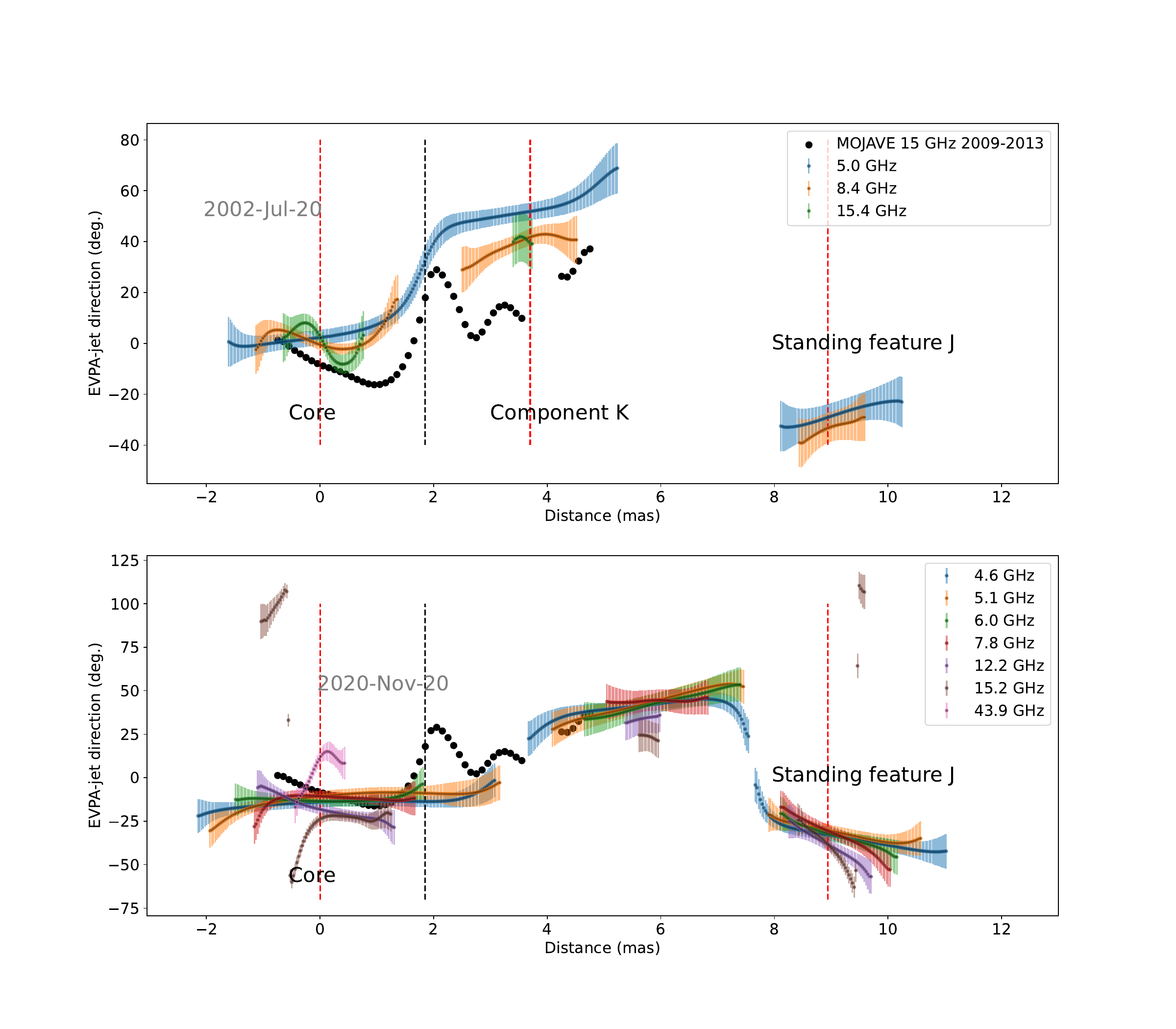}
    \caption{Distributions of EVPA $-$ jet direction along the jet for observations on 2002-Jul-20 and 2020-Nov-20.
    The black scatter points are derived from the MOJAVE stacked results at 15 GHz between 2009-2013.
    The black dashed line divides the source into the central region and the jet region.}
    \label{fig:EVPA-AlongTheJet}
\end{figure*}

\begin{figure}
	\includegraphics[width=0.5\linewidth]{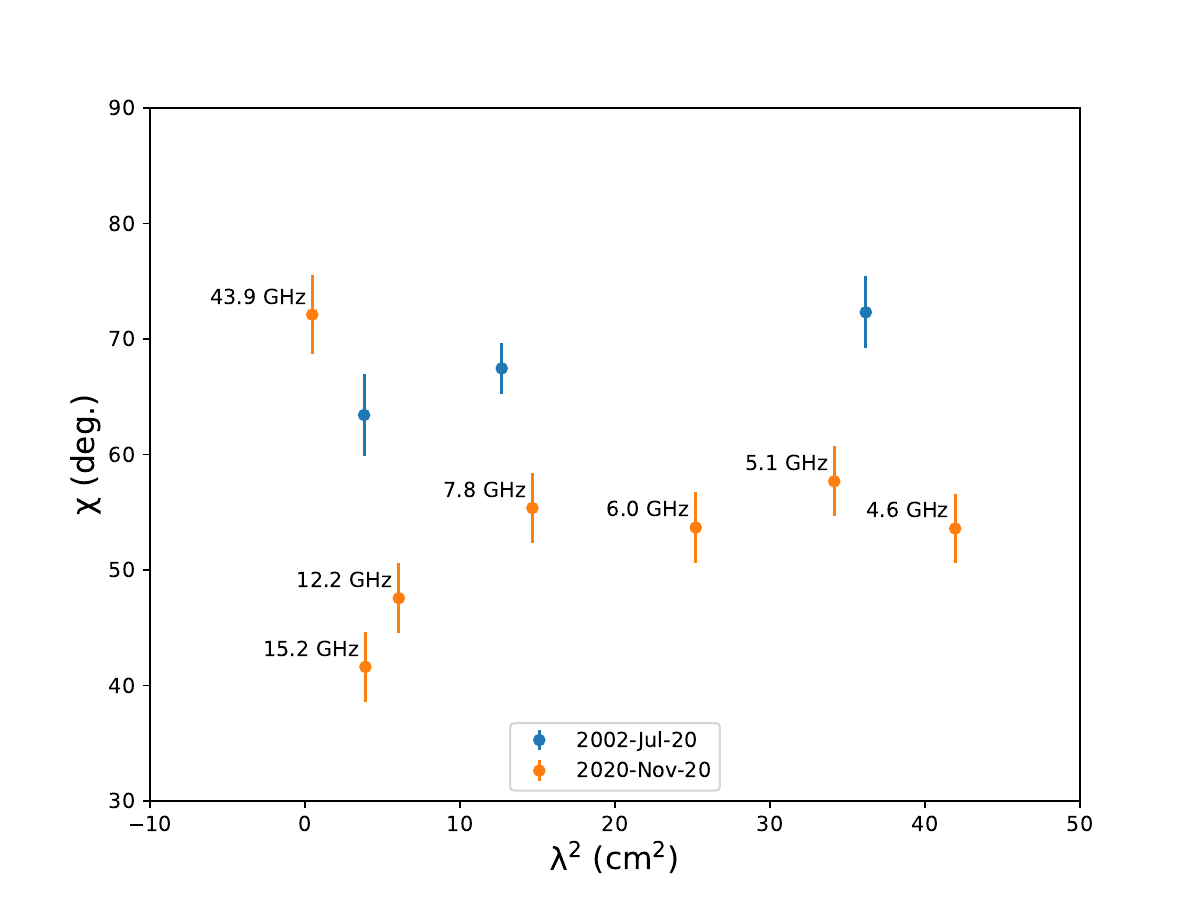}
    \centering
    \caption{A plot of EVPA vs $\lambda^{2}$ for a point sampled from the central region for the multi-frequency observations in 2002 (blue) and 2020 (orange).
    In 2002, EVPA changes linearly with $\lambda^{2}$.
    In 2020, the behavior of EVPA is more complex.}
    \label{fig:1604 EVPA against squared wavelength}
\end{figure}

\begin{figure*}
	\includegraphics[width=\linewidth]{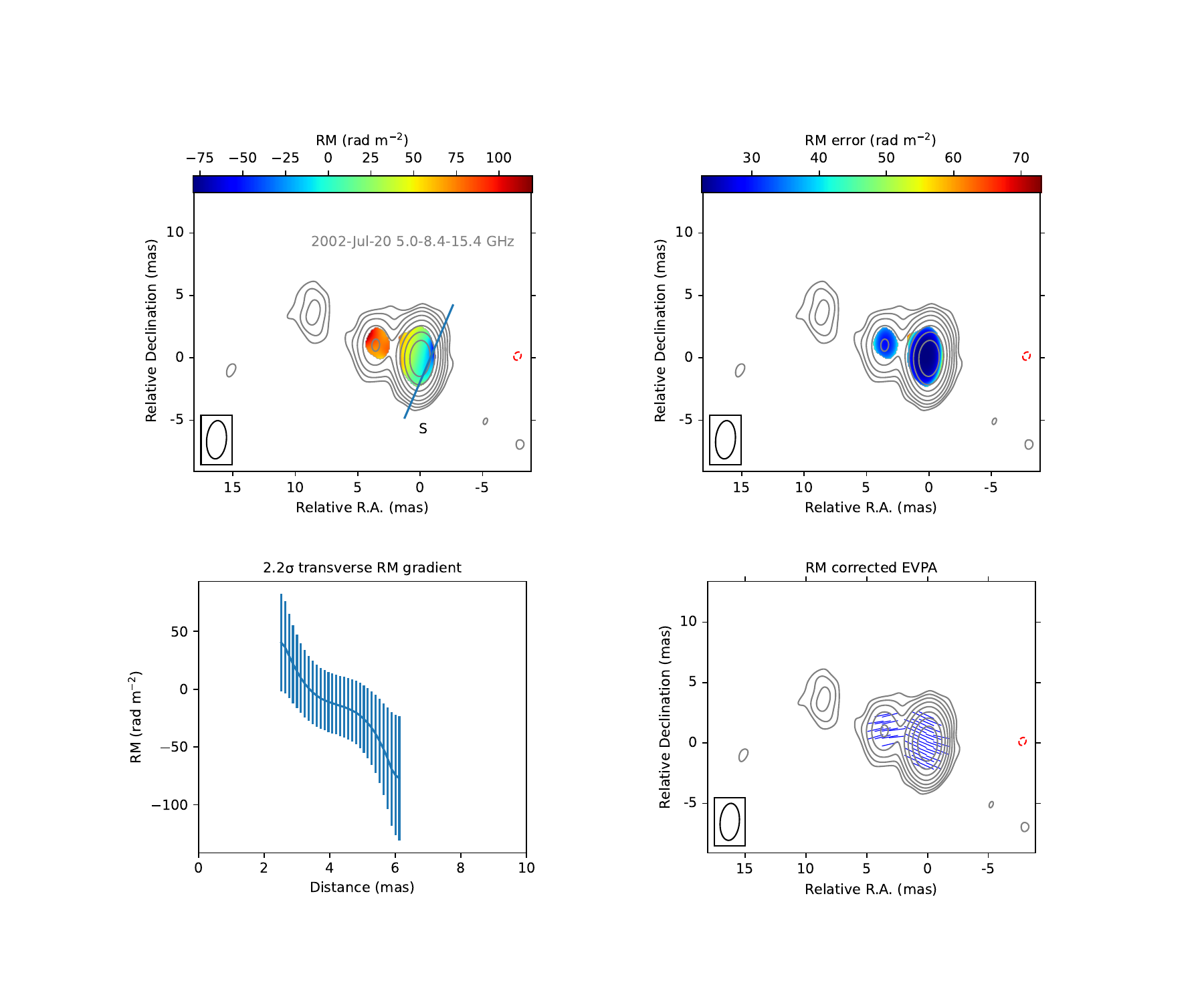}
    \caption{Distributions of RM and RM corrected EVPA for observation on 2002-Jul-20 with a $2\sigma$ transverse RM gradient extracted along the transverse slice (blue solid line) in the RM map.
    The label `S' is the start point of the transverse RM gradient.
    Color is the RM and RM error.
    Contours are the 15.4 GHz total intensity with $3\sigma \times (-1, 1, 2 ,4, 8, 16, 32, 64, 128)$.}
    \label{fig:BH096_RM}
\end{figure*}

\begin{figure*}
	\includegraphics[width=\linewidth]{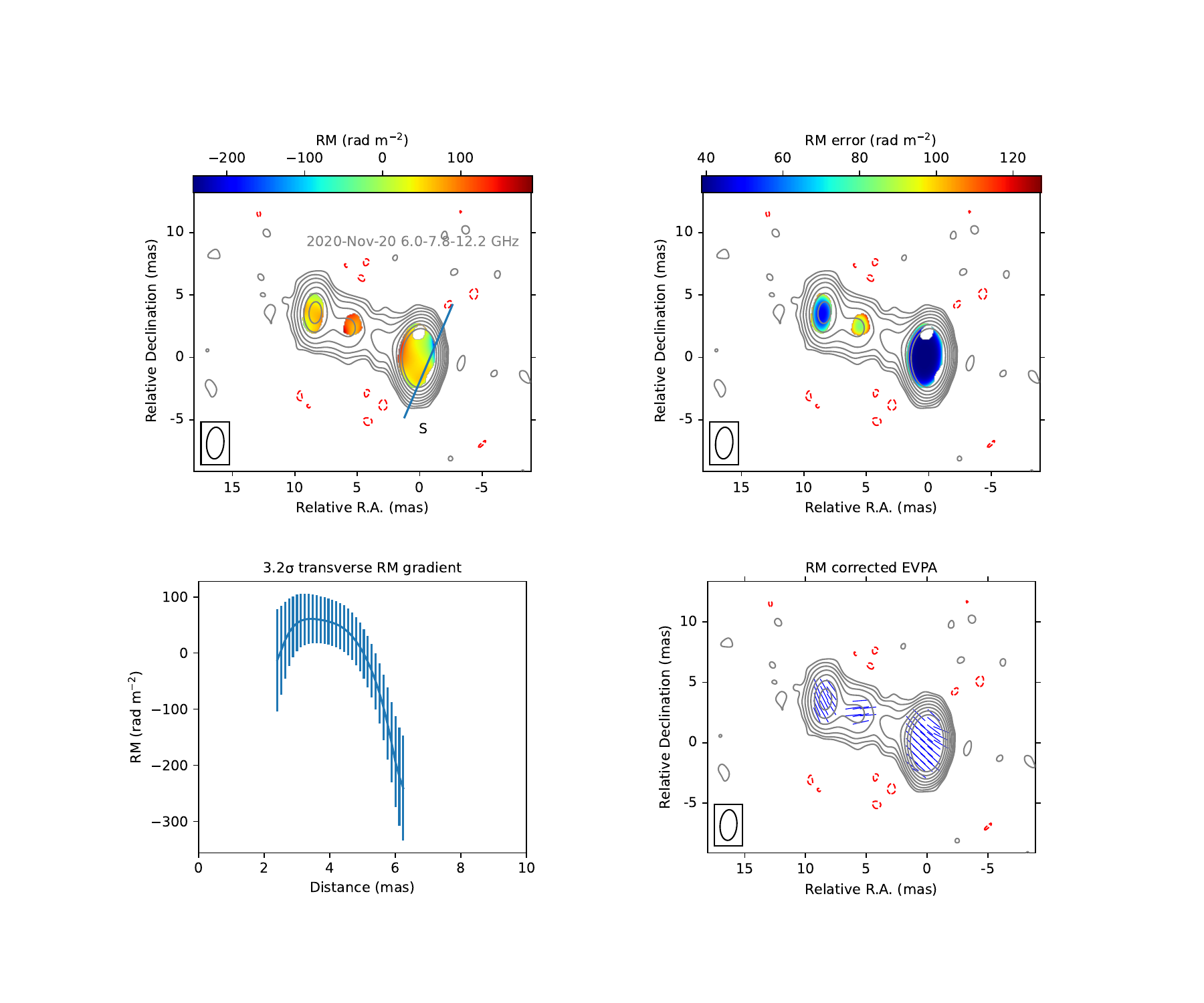}
    \caption{Distributions of RM and RM corrected EVPA at frequency interval 12.2-15.2 GHz for observation on 2020-Nov-20 with a $3\sigma$ transverse RM gradient extracted along the transverse slice (blue solid line) in the RM map.
    The label `S' is the start point of the transverse RM gradient.
    Color is the RM and RM error.
    Contours are the highest frequency total intensity at given frequency intervals with $3\sigma \times (-1, 1, 2 ,4, 8, 16, 32, 64, 128)$.}
    \label{fig:BH228A_RM}
\end{figure*}

\begin{figure*}
	\includegraphics[width=\linewidth]{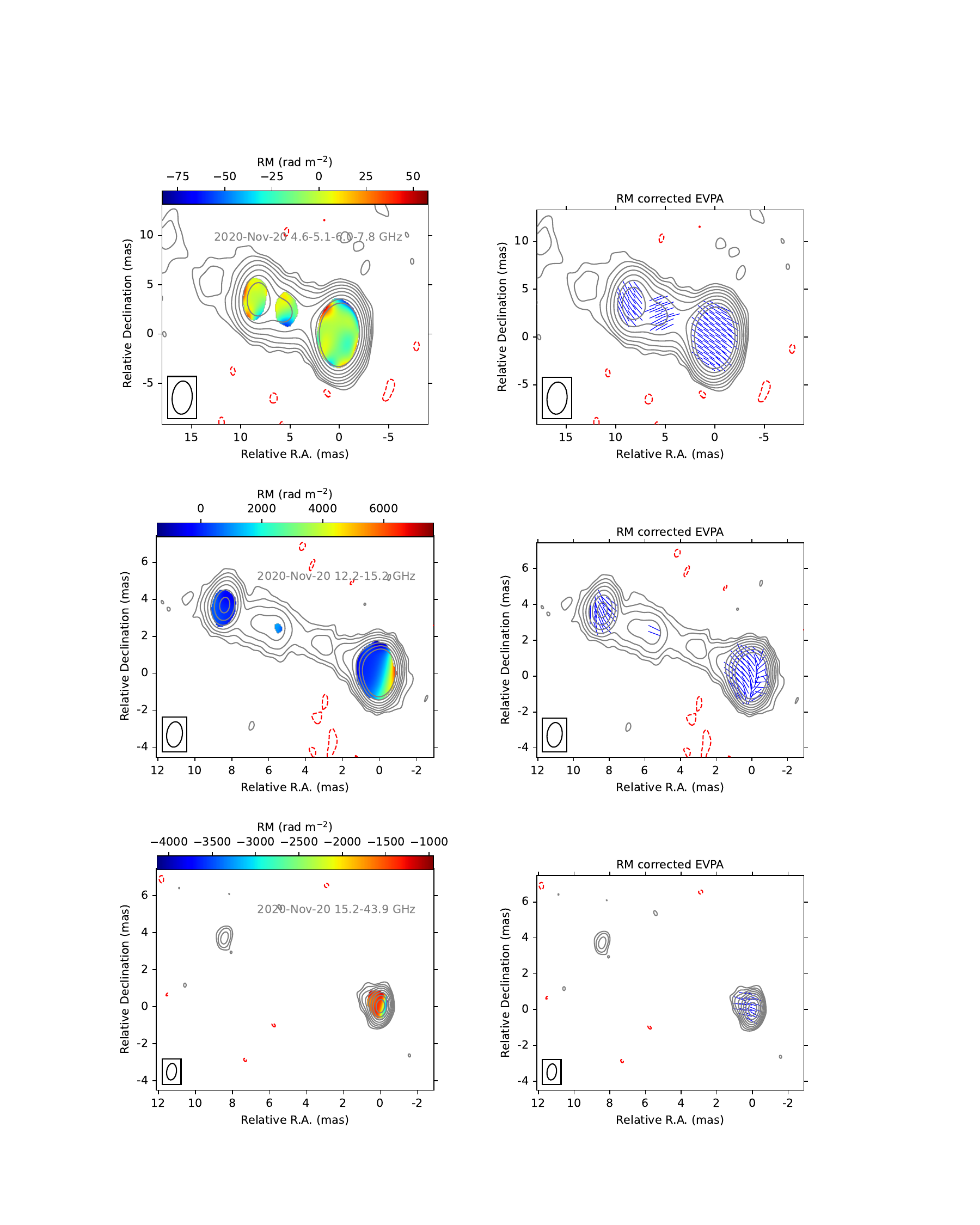}
    \caption{Distributions of RM (left panel) and RM corrected EVPA (right panel) for observation on 2020-Nov-20.
    Frequency intervals are 4.6-5.1-6.0-7.8, 6.0-7.8-12.2, and 15.2-43.9 GHz.
    Color is the RM.
    Contours are the highest frequency total intensity at given frequency intervals with $3\sigma \times (-1, 1, 2 ,4, 8, 16, 32, 64, 128)$.}
    \label{fig:BH228A_RM_V1}
 \end{figure*}

\begin{figure*}
	\includegraphics[width=0.7\linewidth]{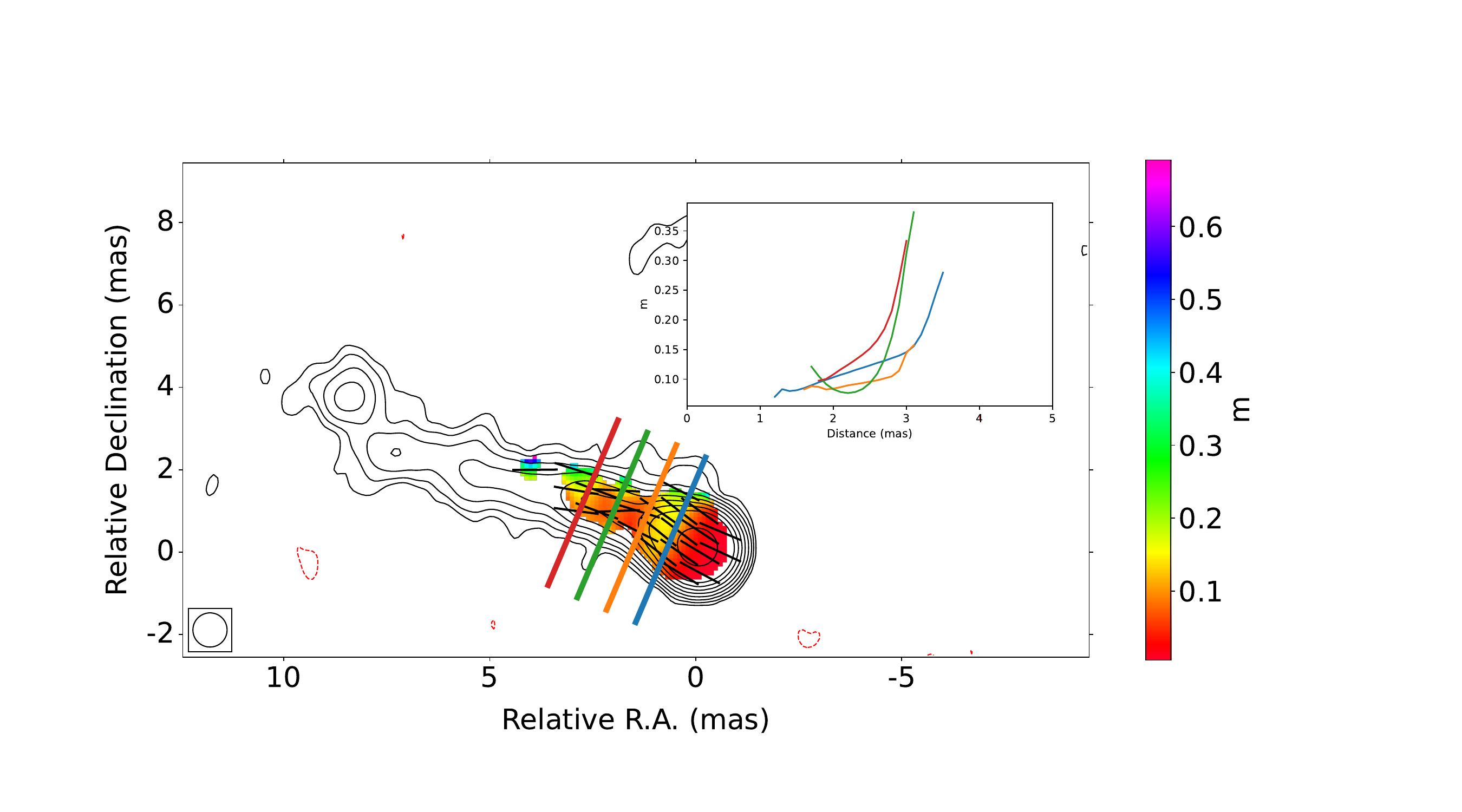}
    \centering
    \caption{MOJAVE stacked image for 1604+159 and transverse slices of fractional polarization.
    Color is fractional polarization.}
    \label{fig:MOJAVE}
\end{figure*}

\begin{table*}
    \centering
    \caption{Observation information}
    \begin{tabular}{c c c c c c c c c}
         \hline
         \hline
         Epoch & Project ID & Array & Frequency & Bandwidth & On Source time & D-term & EVPA & EVPA  \\
           &&&&&&calibrator& calibrator&correction\\
               & & & (GHz) & (MHz) & (min) & & & ($^{\circ}$)\\
         \hline
          & & & 5.0  & & &  & J2253+1608 & 29.8\\
          2002-Jul-20 & BH096 & VLBA & 8.4  & 32  & 56 & OQ208 & J2253+1608 & -41.9\\
          & & & 15.4 &  & & & 3C 279 &-21.9\\
         \hline
         & & & 4.6 & 128  &  & & 3C 286 & 53.3\\
         & & & 5.1 & 128 &  & & 3C 286 & 94.5\\
         & & & 6.0 & 128 &  & & 2200+420 & -31.3\\
         2020-Nov-20 & BH228A & VLBA & 7.8 & 128 & 92 & J1751+0939 & 2200+420 & 59.5\\
         & & & 12.2 & 256 &  & & 2200+420 & 21.1\\
         & & & 15.2 & 256 &  & & 2200+420 & 47.3\\
         & & & 43.9 & 256 &  & & 2200+420 & -7.2\\
         \hline
    \end{tabular}
    \label{tab:observation}
\end{table*}

\begin{table*}
    \centering
    \caption{Image statistics summary.}
    \begin{tabular}{c c c c c c c c}
        \hline
        \hline
         Epoch & Project ID & Frequency & Resolution & $\sigma$ & Integrated  & $I$ peak & Averaged $m$ \\
         & & & & &flux density & & in central region\\
          & & (GHz) & (MilliArc seconds) & (Jy $\rm beam^{-1}$) & (Jy) & (Jy $\rm beam^{-1}$) & (\%) \\
          \hline
          & & 5.0 & $1.64 \times 3.18$ (-5.07) & $1.5 \times 10^{-4}$   & $0.33\pm0.03$ & $0.23\pm0.02$ & $5.3\pm0.5$\\
         2002-Jul-20 & BH096 & 8.4 & $0.99 \times 1.91$ (-3.10) &$1.6 \times 10^{-4}$    & $0.33\pm0.03$ & $0.22\pm0.02$& $6.9\pm0.8$\\
          & & 15.4 & $0.54 \times 1.09$ (-6.94) &$3.1 \times 10^{-4}$   & $0.32\pm0.03$ & $0.22\pm0.02$& $10.1\pm1.5$\\
        \hline
        & & 4.6 & $2.01 \times 3.41$ (-4.58) &$5.5 \times 10^{-5}$  & $0.28\pm0.03$ & $0.22\pm0.02$& $4.3\pm0.5$\\
        & & 5.1 & $1.72 \times 3.15$ (-9.79) &$4.3 \times 10^{-5}$  & $0.28\pm0.03$ & $0.23\pm0.02$& $5.6\pm0.8$\\
        & & 6.0 & $1.40 \times 2.58$ (-4.39)&$8.2 \times 10^{-5}$  &$0.29\pm0.03$ & $0.23\pm0.02$& $5.5\pm0.5$\\
        2020-Nov-20 & BH228A & 7.8 & $1.06 \times 1.99$ (-7.75) &$5.1 \times 10^{-5}$  & $0.28\pm0.03$ & $0.22\pm0.02$& $7.5\pm0.8$\\
        & & 12.2 &$0.85 \times 1.38$ (-6.88) &$7.1 \times 10^{-5}$  & $0.31\pm0.03$ & $0.25\pm0.03$& $7.8\pm1.1$\\
        & & 15.2 &$0.60 \times 1.03$ (-7.59) &$5.0 \times 10^{-5}$  & $0.28\pm0.03$ & $0.21\pm0.02$& $8.1\pm0.8$\\
        & & 43.9 &$0.23 \times 0.51$ (-22.66) &$2.7 \times 10^{-4}$  & $0.21\pm0.02$ & $0.16\pm0.02$& $15.4\pm3.0$\\
        \hline
    \end{tabular}
    \label{tab:Image stat.}
\end{table*}

\begin{table}
    \centering
    \caption{Information of the standing feature J at 15 GHz}
    \begin{tabular}{c c c c}
        \hline
        \hline 
         Obs. date &  Flux density & Distance & Position angle\\
                   &     (mJy)   & (mas) & ($^{\circ}$) \\
        \hline
          2002-Jul-20 & $6.23\pm2.13$ &  $9.33\pm0.03$ &  $66.44\pm0.21$  \\
          2009-Aug-19 & $2.43\pm0.91$ &   $9.20\pm0.07$ &   $66.03\pm0.45$  \\
          2010-Feb-11 & $2.73\pm1.15$ &    $9.22\pm0.08$ &     $67.00\pm0.49$   \\
          2010-Sep-17 & $2.91\pm1.04$ &     $9.26\pm0.08$ &    $66.93\pm0.47$   \\
          2012-Jul-12 & $3.62\pm1.42$ &     $9.18\pm0.13$ &     $66.64\pm0.84$  \\
          2020-Nov-20 & $14.30\pm1.23$ &      $9.22\pm0.01$ &     $66.38\pm0.08$  \\
        \hline
    \end{tabular}
    \label{tab:Component at 9 mas}
\end{table}

\begin{table*}
    \centering
    \caption{4.6 GHz model fitting parameters. Columns: (1) observation date, (2) model fitting method, (3) model component, (4) total flux density, (5) model distance, (6) model position angle, (7) model major axes, (8) model axial ratio, (9) direction of major axes, (10) $\sigma$ after model fitting, (11) reduced $\chi^{2}$ after model fitting.}
    \begin{tabular}{c c c c c c c c c c c}
        \hline
        \hline
         Epoch & Fitting method &Component & $\rm S_{tot}$ & r & $\theta$ & Major axes & Axial ratio & Phi & RMS & Reduced\\
         & && (mJy) & (mas) & ($^{\circ}$) & (mas) && ($^{\circ}$) & (mJy $\rm beam^{-1})$ &  $\chi^{2}$\\
          (1) & (2) & (3) & (4) & (5) & (6) & (7) & (8) & (9) & (10) & (11)\\
         \hline
         & &core&217& 0.012& 88.4& 0.39&0.34& 66.4&&\\ 
         & &...&15 &1.055 & 63.1& 0.35 &1.00 &...&&\\
         2020-Nov-20&elliptical + circular &...&6 &3.081& 63.3 &0.91 &1.00&...&4.97$\times \ 10^{-2}$&1.56\\
         & &...&14& 6.259&     67.5& 2.24& 0.37 &74.2&& \\
         & &J& 25& 9.108& 66.5& 0.77&0.79&29.9&&\\
         \hline
         &&core &202& 0.027&-129.3&0.21&1.00&...&&\\
         &&... & 29& 0.770& 63.0&0.30&1.00&...&&\\
         2020-Nov-20& circular &...&8&3.055& 64.0& 1.44&1.00&...&5.61$\times \ 10^{-2}$&1.57\\
         & &...& 13& 6.277&67.5& 1.67&1.00&...&&\\
         & &J& 25&9.088& 66.5&0.71 &1.00&...&&\\
         \hline
    \end{tabular}
    \label{tab:Model fitting}
\end{table*}

\begin{table*}
    \centering
    \caption{Core shift information. Columns: (1) observation date, (2) component used to derive core shift, (3) frequency pair of core shift, (4) magnitude of core shift, (5) direction of core shift, (6) averaged direction of core shift, (7) core shift index, (8) core shift measure, (9) magnetic field strength assuming equipartition, (10) magnetic field strength without equipartition assumption.}
    \begin{tabular}{c c c c c c c c c c}
        \hline
        \hline
         Epoch &Component & Frequency pair & $\Delta r_{core,\ \nu_{1}\nu_{2}}$ &Direction & $\rm Direction_{ave}$ & $k_{r}$&$\Omega_{r\nu}$ & $B_{1\ pc}^{\rm eq}$ & $B_{1\ pc}^{\rm neq}$\\
         & & (GHz) & (mas) & ($^{\circ})$ & ($^{\circ}$) & &(pc GHz) & (G) & (G)\\
         (1)& (2) & (3) & (4) & (5) & (6) & (7) &(8) & (9) & (10)\\
         \hline
          &K & 5.0-15.4 &$0.181\pm 0.060$ &86.35 &81.03 &$0.77\pm 1.70$ &8.12 &0.325 &0.268 \\
          2002-Jul-20& & 8.4-15.4 &$0.065\pm 0.062$ & 75.71 & & &7.35 & & \\
          &J & 5.0-15.4 &$0.109\pm 0.095$ &-180.74 &-164.85 &$0.56\pm 1.97$ &4.88 &0.208 &0.021 \\
          & & 8.4-15.4 &$0.032\pm 0.075$ &-148.96 & & &3.66 & & \\
        \hline
          & & 4.6-43.9 &$0.114\pm 0.027$ &53.70 & &&3.57  & & \\
          & & 5.1-43.9 &$0.074\pm 0.026$ &57.26 & &&2.61  & & \\
          2020-Nov-20& J & 6.0-43.9 &$0.059\pm 0.029$ &38.63 &36.85 &$0.32\pm 0.11$&2.48  &0.130 &0.003 \\
          & & 7.8-43.9 &$0.030\pm 0.026$ &85.21 & &&1.71  & & \\
          & & 12.2-43.9 &$0.019\pm 0.025$ &30.23 & &&1.98  & & \\
          & & 15.2-43.9 &$0.009\pm 0.025$ & -43.94 & &&1.33  & & \\
         \hline
    \end{tabular}
    \label{tab:core shift}
\end{table*}

\begin{table}
    \centering
    \caption{Core RM observed on 2020-Nov-20}
    \begin{tabular}{c c c c}
        \hline
        \hline 
         Epoch & Frequency interval & RM & a \\
                   &     (GHz)   & ($\rm rad \ m^{-2}$) & \\
        \hline
                   &  $4.6 \sim 7.8$ & $13 \pm 32$ & \\
         2020-Nov-20   &  $6.0 \sim 12.2$ & $53 \pm 52$ & $2.7 \pm\ 0.5$ \\
                   &  $12.2 \sim 15.2$ & $1234 \pm 488$ &  \\
                   &  $15.2 \sim 43.9$ & $1802 \pm 331$ &   \\
        \hline
    \end{tabular}
    \label{tab:Core RM}
\end{table}

\section{results}
\subsection{Total intensity distribution}
%Radio structure of the source 1604+159 at Kpc scale presented by \citet{1998MNRAS.293..257B} had a core and a eastern jet. 
%Our observation detected double extended jets (Fig.~\ref{fig:VLA}) roughly along the east-west direction at kpc scale in which two components contribute the prominent emission.
%This indicates that source may have a large view angle.
%The two components both have shown knot structures and their distance to the core are close.
%The eastern one has a distance of 11.0 arcsec with position angel of $95.7^\circ$ from the core, and the western one has a distance of 11.5 arcsec with position angle of $-91.8^\circ$.
%Model-fitting results have shown that the eastern knot has flux $\sim$ 2.5 times higher than the western one.

Fig.~\ref{fig:BH096 pol} and Fig.~\ref{fig:BH228A pol} show the stokes $I$ distributions for the source 1604+159 at pc scale.
There is a core-jet structure.
The jet propagates to the direction of $\sim 66^{\circ}$ north by east.
In 2002, the jet structure extends to a distance of $\sim 15$ mas from the core (5.0 GHz in Fig.~\ref{fig:BH096 pol}).
With higher sensitivity, in 2020, the jet extending to a distance of $\sim 25$ mas from the core is detected (4.6 and 5.1 GHz in Fig.~\ref{fig:BH228A pol}).
The total intensity map at 43.9 GHz (Fig.~\ref{fig:BH228A pol}) shows the core and a jet feature (labeled as J) located at $\sim 9$ mas from the core.
To our knowledge, this is the first time the quasar 1604+159 is detected at 44 GHz.
The information on the images is listed in Table~\ref{tab:Image stat.}.

%-------------------------------

%In order to describe the shape of the jet, we fitted circular or elliptical Gaussian components for the 15 GHz data in Difmap and then calculated cubic spline lines through these components for both epochs separately.
%The resultant map is plotted on Fig. \ref{fig:Ridge Line}.
%The reason of using the combination of circular and elliptical Gaussian components for the model fitting is that it behaves better than using all the circular Gaussian components, and the Root Mean Square (RMS) of the residual maps after fitting is close to that after clean.

%---------------------------------

%+++++++

In order to probe whether the shape of the jet varies with time, we collected our 15 GHz data and the Mojave 15 GHz data from 2009 $\sim$ 2012 \citep{2018ApJS..234...12L} and used the method described in \citet{2015ApJ...803....3C,2017MNRAS.468.4992P} to derive the jet streamline.
This method starts from the core and finds the ridge line points at the same intervals (0.6 mas in this study), which were then connected with a cubic spline.
At each ridge line point, the integral of the intensity along a circular arc across the jet, which has the core as its center, is equal on the two sides of the arc. 
As the method may be affected by the restored elliptical beam \citep{2017MNRAS.468.4992P}, we restore the median of circular beams derived from the naturally weighted elliptical beams ($\rm d=\sqrt{major \ axis \times minor \ axis}$) for all the 15 GHz images.
During the calculation, we blanked pixels with a total intensity  $< \ 6\sigma$.

The resultant streamlines are plotted in the Fig. \ref{fig:Ridge Line}.
We found the jet propagates along the direction between the core and a feature J.
%++++++++++++
%--------------
%Combining with the stacked results at 15 GHz from 2009 to 2013 \citep{2023MNRAS.520.6053P}, we find the shape of the jet changes with time. 
%The jet in 2020 has a slight C-like shape, then becomes slight S-like shape during 2009 $\sim$ 2013 \citep{2023MNRAS.520.6053P}, then becomes more wobbling in 2020.
%-------------
We also found the jet feature J exists at all frequencies in the two epochs, the MOJAVE results from $2009\sim2012$ \citep{2018ApJS..234...12L}, and the VLBA Imaging and Polarimetry Survey (VIPS) results \citep{2007ApJ...658..203H}.
We believe this is a standing feature detected in most of the observations spanning about 18 years from 2002 to 2020.
The information on the standing feature at 15 GHz is listed in Table~\ref{tab:Component at 9 mas}.

%+++++++++++++++++
Interestingly, in the content between the core and the $\sim$ 6 mas place, there is a hint that the jet rotates clockwise with time from south shown by the jet in 2002 (blue solid line) to north in 2020 (pink solid line).
The physical origins behind this phenomenon include the existence of a helical magnetic field as shown by \citet{2021ApJ...906..105C}, jet procession like M87 \citep{2023Natur.621..711C}, interaction with the medium surrounding the jet.
To note, further long-term observations are needed to confirm whether the rotation does exist and to probe the physical origin.
%++++++++++++++++++

Although the jet shape slightly changes with time, it goes through the region where standing feature J is located.
Besides, maps at the C band in 2020 (Fig.~\ref{fig:BH228A pol})  show the eastern jet of 1604+159 continues to propagate through the feature without changing its direction.
%To get the spectral variation along frequency and time, we performed model-fitting for this feature for the data of BH096,  BH228A, and 5 epochs of MOJAVE 2 cm +++.
%Fig.~+++ shows the flux density against frequency and time.
%As can be seen, this feature do not follow a simple power law.
Thus, we took a straight line through the core and the standing feature J as its jet ridge line (Fig. \ref{fig:Ridge Line}) for the following spectral and polarization analysis.

%According to  \citet{2002ApJ...568...99H,2009AJ....138.1874L,2019ApJ...874...43L},
%the flux uncertainties are about of 5 percent of the flux densities. 
%The uncertainties of centroid positions of
%the components  typically are about 1/5 of the FWHM naturally-weighted image restoring beam dimensions.
%For the two bright compact knots and the core, the positional errors are smaller by approximately a factor of two.

\subsection{Core shift and magnetic field strength}
As one of the most important properties of a magnetic field, magnetic field strength at 1 pc distance from the central engine could be estimated from core shift measure using our multi-frequency data.
According to \citet{1979ApJ...232...34B}, the core is located at a region where the optical depth, $\tau$, roughly is one.
Due to synchrotron self-absorption, the core at different frequencies is observed in different regions.
Assuming a freely expanding, supersonic, narrow, conical jet, the distance of the core from the central engine follows a power law, $r_{core}$ $\propto$ $\nu _{obs} ^{-1/k_{r}}$, where $k_{r}$ = 1 corresponds to the energy equipartition case.

To measure the core shift, we fitted circular or elliptical Gaussian components in Difmap and then obtained the fitted information for components to be used for both epochs, separately.

For data in 2002, feature J and the component (labeled as K as shown in Fig. \ref{fig:BH096 pol}) roughly at 3.7 mas from the core were used.
For data in 2020, the feature J is the only component existing at all 7 frequencies and was then used.

The reason for using the combination of circular and elliptical Gaussian components for the model fitting is that it behaves better than using all the circular Gaussian components, and the Root Mean Square (RMS) of the residual maps after fitting is closer to that after clean.

Table~\ref{tab:Model fitting} lists an example of the fitting parameters of components for 4.6 GHz data in 2020. It also provides information (e.g. RMS, $\chi^{2}$) for comparison between using the combination of circular and elliptical Gaussian components for the model fitting and using all the circular Gaussian components.

The uncertainties of the fitting parameters were obtained following the approach of \citet{2008AJ....136..159L}, which takes into account the effect of signal-to-noise (S/N).
For data at\textbf{ }the C band, extended diffuse emissions beyond the standing feature J were cleaned to decrease the RMS of the model images.

Core shift vector ($|\textbf{CS}|=\Delta r_{core,\ \nu_{1}\nu_{2}}$) was measured using the same method of \citet{2012A&A...545A.113P}  for frequency pairs including the highest frequency and each lower frequency data for the two observations, separately.
This method derives the core shift from \textbf{CS} = \textbf{IS} - \textbf{OS}, where \textbf{IS} is the difference of the image shift (difference of compact component coordinates of images for a given frequency pair), and OS is the difference of the core coordinates.
The corresponding uncertainties were derived by using the method described in the study of \citet{2021A&A...652A..14C,2023A&A...672A.130C}.
In our study, the uncertainty of the core shift is estimated as
\begin{equation}
    \label{uncertainty of core shift}
    \sigma_{\Delta r_{core,\ \nu_{1}\nu_{2}}} = 
    \sqrt{\sigma_{IS,\ \nu_{1}\nu_{2}}^{2}+
    \sigma_{OS,\ \nu_{1}\nu_{2}}^{2}},
\end{equation}
where $\sigma_{IS,\ \nu_{1}\nu_{2}}$, which equals  $\sqrt{\sigma_{r_{J(K),\ \nu_{1}}}^{2}+\sigma_{r_{J(K),\ \nu_{2}}}^{2}}$, is the uncertainty of aligning images arising from the position uncertainties of the used compact component between frequency pair $\nu_{1}$ and $\nu_{2}$, and $\sigma_{OS,\ \nu_{1}\nu_{2}}$ is the sum (in quadrature) of the position uncertainties of the core, $\sqrt{\sigma_{r_{core,\ \nu_{1}}}^{2}+\sigma_{r_{core, \ \nu_{2}}}^{2}}$. The resultant core shift information is listed in Table~\ref{tab:core shift}.

In 2002, the core shift derived from components J and K show similar magnitude but different directions.
The directions of core shift derived from K ($\sim$ 15 $^{\circ}$ of separation from the jet direction) are\textbf{ }closer to the jet direction than that from J ($\sim$ 129 $^{\circ}$ of separation from the jet direction).  
We noted that the emission of component K is stronger than that of component J, resulting in smaller uncertainties of position.
Also, the core shift was derived using vector computing and is small compared with the distances of the two components from the image centers.
Thus, the derived core shift (especially its direction) is more easily affected by the larger uncertainty of position and further distance of the component used.

With the information on core shift vectors, we obtained  absolute projected values and uncertainties (on the mean direction of the core shift vectors) of the core shift and then fitted these values with a power law curve,
\begin{equation}
    \label{fitting of core shift}
    \Delta r_{core} = 
    a(\nu^{-1/k_{r}}-\nu_{ref}^{-1/k_{r}}),
\end{equation}
where $\Delta r_{core}$ is the core shift in mas, $\nu$ is the observed frequency in GHz, $\nu_{ref}$ is the highest observed frequency for a given observation (15.4 GHz for the observation in 2002 and 43.9 GHz for the observation in 2020), a is a fitting coefficient, and $k_{r}$ is the power law index of core shift. The fitted results are plotted in Fig. \ref{fig:1604 coreshift}.

$k_{r}$ in 2002 has large uncertainties, compared with that in 2020.
This may be due to the absence of more samples from other frequencies especially lower frequencies (e.g. L-band).
As studied by \citet{2011A&A...532A..38S}, core shift effect is most obvious at lower frequencies and the uncertainty of $k_{r}$ would be significantly decreased by adding L-band data.

%The measured values of $k_{r}$ for 2002 and 2020 show hint of variation with time.
In 2002, as discussed above, due to the stronger emission and smaller distance to the core of the component K, the $k_{r}$ derived would be more convincing.
In 2002,  $k_{r}$ is close to 1, the equipartition case.
However, $k_{r}$ measured in 2020 shows a larger deviation.
The value of $k_{r}$ (lower  than 1) and its change with time is very\textbf{ }similar to the blazar 3C 454.3 studied by \citet{2023A&A...672A.130C} with wide multi-frequency and multi-epoch observations, who found the value of $k_{r}$ for 3C 454.3 varies between $\sim$ 1, and values lower than 1.

%blending effect

%magnetic field strength
Following \citet{1998A&A...330...79L,2012A&A...545A.113P,2014Natur.510..126Z,2015MNRAS.451..927Z}, we delivered the magnetic field strength at 1 pc using both equipartition and non-equipartition methods for the two epochs.
Both assumptions need information on the half-opening angle $\varphi$, the Doppler factor $\delta$, and the viewing angle $\theta$.
To calculate $\theta$, the Doppler factor $\delta$ has to be known. For the source 1604+159, common estimations of $\delta$ including from combining X-ray observations with Very Long Baseline Interferometer (VLBI) component fluxes \citep{1993ApJ...407...65G} and brightness temperature of the source ($T_{b,\ obs}$) \citep{1994ApJ...426...51R} need VLBI frequency observed at the turnover frequency and equipartition assumption, which is not the case for the source 1604+159.
In 2002, our VLBI observation did not show the turnover, and in 2020, the derived $k_{r}$ suggests deviations from the equipartition.
As the source 1604+159 shows low radio variability \citep{2014A&A...572A..59M,2022AstBu..77..361S}, the variability method \citep{2005AJ....130.1418J,2009A&A...494..527H} may also not give good estimation of the Doppler factor $\delta$ and thus the viewing angle $\theta$.
Therefore, we used the same assumption of \citet{2012A&A...545A.113P} to estimate the magnetic field strength.
As suggested by \citet{2014Natur.510..126Z},  when the viewing angle could not be estimated accurately, the jet could be assumed by viewed close to their critical angle, $\theta \approx \Gamma^{-1}$ \citep{2012A&A...545A.113P} for the estimation of magnetic field strength.
In that case, $\delta\approx \Gamma$ and $\Gamma=\sqrt{1+\beta_{\rm app}^{2}}$.
We also assumed the half-opening\textbf{ }angle $\varphi=0.13\Gamma^{-1}$ \citep{2012A&A...545A.113P}.

If equipartition holds, the magnetic field strength could be estimated through the following equation
\begin{equation}
    \label{B-eq}
    B_{1\ \rm pc} \approx
    0.025(\frac{\Omega_{r\nu}^{3}(1+z)^{3}}{\varphi\delta^{2}\rm sin^{2}\theta})^{\frac{1}{4}}
\end{equation}
 \citep{1998A&A...330...79L}, where $\varphi$ is the half jet opening angle, $\theta$ is the viewing angle, $\delta$ is the Doppler factor, and $\Omega_{r\nu}$ is the core shift measure defined as 
\begin{equation}
    \label{core shift measure}
    \Omega_{r\nu} =
    4.85\times10^{-9}
    \frac{\Delta r_{core,\ \nu_{1}\nu_{2}}D_{L}}{(1+z)^{2}}\times
    \frac{\nu_{1}\nu_{2}}{\nu_{2}-\nu_{1}} \rm pc \ GHz,
\end{equation}
where $D_{L}$  is the luminosity distance in parsec.

For the non-equipartition case, the magnetic field strength could be calculated by the equation
\begin{equation}
    \label{B-neq}
    B_{1\ \rm pc} \approx
    \frac{3.35\times10^{-11}D_{L}\delta\Delta r_{core,\ \nu_{1}\nu_{2}}^{5} \rm tan^{2}\varphi}
    {(\nu_{1}^{-1}-\nu_{2}^{-1})^{5}[(1+z)\rm sin\theta]^{3} \textit{F}_{\nu}^{2}}
\end{equation}
\citep{2015MNRAS.451..927Z}.
The obtained magnetic strength values are also listed in  Table~\ref{tab:core shift}.

We found in 2002, with core shift derived from component K, both the values of the $B_{1\ \rm pc}$ measured by using the equipartition and non-equipartition methods are close to each other in magnitude.
However, in 2020, the non-equipartition case shows the strength of $B_{1\ \rm pc}$ two orders of magnitude lower than the equipartition case, consistent with the magnetic field study of part of 59 blazars for the two cases by \citet{2015MNRAS.451..927Z}. 
The $B_{1\ \rm pc}$ derived for 1604+159 also shows change with time.
A more comprehensive discussion will be presented in Section 4.

\subsection{Spectral index}
To avoid spurious gradients, images need to be aligned before the construction of spectral index maps due to loss of absolute position after self-calibration for each frequency.
Two main methods can be used to align images. 
One is using bright and compact components in the jet assuming their optically thin nature \citep[e.g.][]{1998A&A...330...79L,2001ApJ...550..160M,2008A&A...483..759K,2011A&A...532A..38S,2012AJ....144..105H}. 
The other is two-dimensional (2D) cross-correlation to obtain its best core-shift for each frequency through calculation of the largest correlation coefficient of optically thin jet for each frequency \citep{2000ApJ...530..233W,2008MNRAS.386..619C}.
As the jet of 1604+159 at pc scale has several bright and compact components detected (Fig.~\ref{fig:BH096 pol} and ~\ref{fig:BH228A pol}), the former aligning method was adopted.

%determination of frequency intervals
Source 1604+159 is a LSP Quasar with a spectral peak located at a frequency  $<10^{14}$ Hz.
Compared with the optically thin nature of the jet region, the spectra in the central region may behave more complexly.
To determine the frequency intervals for spectral index maps, 
%we plotted the total flux density and peak intensity of the core obtained from the model fitting against frequency in Fig++++++.
we re-produced $I$ distributions for each frequency with\textbf{ }the same pixel size, shifted image centers to the centroid of the standing feature J, and converted the images to the beam size of the lowest frequency image for both epochs.
Then 5 points (roughly the same interval of $\sim$ 0.6 mas) were sampled along the jet direction from the central region, shown in Fig.~\ref{fig:1604 I vs frequency}.

As seen in Fig.~\ref{fig:1604 I vs frequency}, in 2020, the total intensity of 4.6-5.1-6.0 GHz and 12.2-15.2-43.9 GHz follow a power law for the 5 sampled points in the central region.
However, the total intensity at 7.8 GHz shows more complex behavior.
It drops earlier than that at 12.2 GHz and higher frequencies (as shown from (1)-(5) in Fig.~\ref{fig:1604 I vs frequency}) when being away from the jet base along the jet direction. 
Based on that, spectral index maps were constructed for the following frequency intervals:
4.6-5.1-6.0, 6.0-7.8, 7.8-12.2, and 12.2-15.2-43.9 GHz in 2020;
5.0-8.4 and 8.4-15.4 GHz in 2002.

Take 4.6-5.1-6.0 GHz data in 2020 as an example:
The (u,v)-coverage was first flagged to the same range in which the minimum UV distance is the smallest value of the highest frequency and the maximum is the largest one of the lowest frequency.
Then, maps were restored to the lowest frequency beam size.
Last, the map center of each frequency was shifted to the position of the standing feature J obtained from model fitting to align the images, which was also used by \citep{2022MNRAS.510.1480K}.
Other frequency intervals were treated with the same process.

To derive the spectral index and its uncertainty pixel by pixel, we used the same method of \citet{2014AJ....147..143H} .
More specifically, the spectral index distribution ($S \propto \nu^{\alpha}$) of each frequency interval was obtained by fitting a power-law to the total intensity distribution of each frequency in the interval.
When fitting the spectral index, we added in quadrature a calibration uncertainty of 0.1.
Pixels were clipped when the total intensity level was less than $3\sigma_\nu$ at the corresponding frequency, where
\begin{equation}
    \rm \sigma_{\nu} = \sqrt{\sigma^2_{rms} + (1.5\sigma_{rms})^2} \approx 1.8\sigma_{rms},
\end{equation}
where $\sigma_{\rm rms}$ is the thermal noise of the image taken at a corner of each map.
We also blanked pixels with uncertainty larger than 1.
The resultant map is shown in Fig.~\ref{fig:1604 Spectral index}.

To analyze the spectral distribution along the source,  we extracted the spectral index values along the ridge line, as shown in Fig.~\ref{fig:SPIX along the jet}.
We divided the emission region into two regions based on the polarization distribution in 2002 (Fig.~\ref{fig:BH096 pol}): the central region including the core and upstream of the jet, and the jet region.

%In the central region, the spectral index experiences a smooth change from a slight inverted value  $\sim 0.25$ to flat value $\sim 0$ to steep value ($< 0$).
In the central region, the core in general has flat spectra.
For frequency intervals 8.4-15.4 GHz in 2002 and 7.8-12.2 GHz in 2020, the spectral index experiences a smooth change from a slight inverted value ($>0$) to a flat value ($\sim 0$) to a steep value ($< 0$).
This smooth transition is due to the convolution with the beam, but the change is intrinsic to the source, as shown in the simulations by \citet{2014AJ....147..143H}, indicating that the central region of the source 1604+159 experiences a spectral change from synchrotron self-absorbed to flat to optically thin.
This change with distance also reflects the opacity change with distance in the core.

For the jet region, \citet{2014AJ....147..143H} measured spectral index values from the jet ridge lines.
They found quasars have a mean spectral index of $-1.09 \pm 0.04$.
%As can be seen in Fig.~\ref{fig:SPIX along the jet}, in general, spectral index values in the jet of the quasar 1604+159 has larger spectral index values.
%We noted that, in these two epochs, regions with bright, compact jet features have flatter spectral including the standing feature J.
Compared with this mean value, we noted that, in the two epochs, regions with bright, compact jet features have larger spectral values including the component K and the standing feature J.
Assuming a homogeneous synchrotron for these regions, the spectral index $\alpha = \frac{1-p}{2}$, where p is the index of the energy distribution of electrons $N(E) = N_{0}E^{-p}$, where E is the electron energy.
The increased spectral indices indicate the increase of electron energy in the compact features, which suggests that electrons in the regions are accelerated.
This supports the existence of shocks in these flatter regions.
The standing feature J detected in most of the observations from 2002 to 2020 provides another evidence of the existence of shock in this region.

\subsection{Polarization}
\subsubsection{Linear polarization distribution}
Fig.~\ref{fig:BH096 pol} and Fig.~\ref{fig:BH228A pol} also show linear polarization and fractional linear polarization ($m$) distributions for 1604+159.
The polarization pixel was clipped when its intensity was below $3\sigma_{p}$ for each frequency, where $\sigma_{p}$ is derived according to the polarization uncertainty analysis of  \citet{2012AJ....144..105H}.

The central region contains polarized emissions from the core and the upstream of the jet, the averaged $m$ is 5.3$\%$ at 5.0 GHz, 6.9$\%$ at 8.4 GHz, and 10.2$\%$ at 15.4 GHz in 2002, indicating the depolarization effect due to reasons including Faraday Rotation and opacity effects.
In 2020, the averaged $m$ shows a similar depolarization effect, as shown in Table~\ref{tab:Image stat.}.

In the jet, polarized emissions are centered in areas with bright, compact features.
We found the fractional polarization is higher than that in the central region.
More and more evidences of increase of fractional polarization with distance from the jet base indicate that magnetic field becomes more ordered along the jet\citep{1993ApJ...416..519C,2000ApJ...541...66L,2001ApJ...562..208L,2005AJ....130.1389L,2023MNRAS.520.6053P}, which may due to reasons including small Faraday rotation \citep{2012AJ....144..105H} and spectral ageing effect \citep{1962SvA.....6..317K}.

In the core, observations at different frequencies reveal emissions from different regions, with higher frequencies corresponding to regions closer to the jet base.
The surrounding medium of the core at different frequencies may also have diverse properties including spectral index and Rotation Measure, which may lead to complex distribution of the observed polarization.
To analyze the observed EVPA along the jet, we obtained the distributions of EVPA minus jet direction along the jet ridge line for the two observations, as shown in Fig.~\ref{fig:EVPA-AlongTheJet}.

For the central region, \citet{2023MNRAS.520.6053P} has analyzed the EVPA orientation to the jet direction along the ridge line for blazars.
They found that for the core of quasars the $\rm |EVPA - jet\ PA|$ has a clear predominance of a peak near $90^{\circ}$ and a weaker secondary peak at values close to $0^{\circ}$, in contrast to BL Lacs which show more predominant EVPAs aligned with the jet direction.
As seen in Fig.~\ref{fig:EVPA-AlongTheJet}, the core of 1604+159 has a simple consistent polarization direction basically along the jet except for the 15 GHz in 2020.
In 2020, a complex variation of the EVPA with distance appears in the 15 GHz core, which experiences a rapid change from perpendicular ($\sim 90^{\circ}$) to parallel ($\sim -25^{\circ}$) to the jet direction starting at an area close to the jet base.
In contrast, in 2002, the EVPA of the 15 GHz core in general orients toward the jet direction.
This suggests that the polarization of the 15 GHz core evolves with time, confirming the results of \citep{2018ApJS..234...12L}  which have shown a similar large change of EVPA with time in the area close to the jet  base of 1604+159.
%We find at 15 GHz the EVPA in the core shows clearly different for the two epochs.
%In 2002, the EVPA of the 15 GHz core in general orients towards the jet direction with small variation.
%However, in 2020, large variation of the EVPA appears in the 15 GHz core, which experiences a smooth change from perpendicular to parallel to the jet direction starting at area close to the jet base.

In the jet region, \citet{2023MNRAS.520.6053P} also found that the EVPAs in quasars tend to be preferentially transverse to the jet in the outer ridge line regions.
For the jet of 1604+159, we did not detect the EVPA transverse to the jet but had a large separation of $\sim \pm 40^{\circ}$ from the jet direction, where positive and negative signs are counterclockwise and clockwise, respectively.
Compared with that of  \citep{2018ApJS..234...12L} and \citet{2023MNRAS.520.6053P}, we also found the observed EVPA changes with time in the jet.
The origin of the change of EVPA in the central region and the jet region will be discussed in Section 4.

\subsubsection{Rotation Measure}
%Observation at different frequencies reveals emission from different scales. 
Our broad-band frequency ($4.6 \sim 43.9$  GHz) data in 2020 allows us to probe RM at different scales from different frequency intervals.
To determine the frequency intervals, we imaged polarization distribution for each frequency observed in 2020 with the same pixel size, shifted image centers to the centroid of the standing feature J, and converted the images to the beam size of the 4.6 GHz image.
Then a point was sampled from the central region in the images of the seven frequencies, and the EVPA distribution against squared wavelength was obtained for the point as shown in Fig. \ref{fig:1604 EVPA against squared wavelength}.
Observation in 2002 was done in the same process.

As can be seen from Fig. \ref{fig:1604 EVPA against squared wavelength}, in 2020, the EVPA stays stable at the C band and decreased at 12.2 GHz following a non-linear behavior, 
with larger EVPA rotation at larger frequencies, leading to a possible relation between RM and frequency, which will be analyzed and discussed in Section 4.
In 2002, the EVPA seems to show a more simple rotation.
This may be due to the absence of observation at other frequencies or the existence of a simple external medium.
Given its frequency range and sample number (5.0, 8.4, and 15.4 GHz), this simple rotation of EVPA could be fitted by a straight line.
The foreground RM arising by our Milky Way for the source 1604+159 is negligible because it is located in one of the ’holes’ where the amplitude of the median RM is consistently below 1 $\textit{\rm rad  m}^{-2}$ \citep{2009ApJ...702.1230T}.
Thus, the rotation of EVPA shown in Fig. \ref{fig:1604 EVPA against squared wavelength} probably reflects the intrinsic physical condition in the central region of 1604+159 and its environments.

%The stable EVPA at C band in 2020 suggests the source do not suffer obviously from external Faraday screen, and the rotation of EVPA at high frequency comes from internal.
%On the contrary, in 2002 the rotation of EVPA may mainly comes from a external Faraday screen.

We then selected four frequency intervals for the observation in 2020 to construct RM distributions which include 4.6-5.1-6.0-7.8 GHz, 6.0-7.8-12.2 GHz, 12.2-15.2 GHz, and 15.2-43.9 GHz.
RM distribution in 2002 was also constructed.

The multi-frequency data allows us to access the Rotation Measure (RM) according to the equation \ref{RM}.
Pixel was blanked when the chi-squared ($\rm \chi^2$) exceeded 3.84 (5.99), corresponding to a 95\% confidence limit for 1 (2) degree of freedom. The $\chi^2$ is from the standard equation:
\begin{equation}
  	 \chi^2 = \sum \frac{( O_i - E_i)^2}{\sigma_i^2},
\end{equation}
where $O_i$  are the observed values, $E_i$ are the expected values of the fitting model, and $\sigma_i$ are the EVPAs variances \citep[e.g.][]{1992nrfa.book.....P,2010arXiv1008.4686H}.

When constructing RM distributions for the frequency intervals 12.2-15.2 GHz and 15.2-43.9 GHz, we minimized the modulus of RM by shifting the EVPA at one frequency by $n\pi$ to solve the EVPA ambiguity problem \citep{2022MNRAS.510.1480K}.
The resulting RM distributions are plotted in Fig.~\ref{fig:BH096_RM}, Fig.~\ref{fig:BH228A_RM}, and Fig.~\ref{fig:BH228A_RM_V1}.

RM serves as an important tool to probe the \textbf{\textit{B}} field since it carries information on the projected \textbf{\textit{B}} field on the LoS, as shown by equation \ref{RM}.
Detection of transverse RM gradients with significance of $3\sigma$ and above provides strong evidence for the existence of helical \textbf{\textit{B}} field \citep{2012AJ....144..105H,2018A&A...612A..67G}.
The majority of transverse RM gradients were found at central regions including cores and upstream of the jets of blazers (e.g. 0059+581, 1124-186, 1908-201 \citep{2018A&A...612A..67G}).
For the observation in 2002, we detected a transverse RM gradient with $2\sigma$ significance at the core region of 1604+159, as shown in Fig. \ref{fig:BH096_RM}.
This corresponds to a confidence level  $> 90 \%$ \citep{2012AJ....144..105H}.
In 2020, at almost the same place, we detected a transverse RM gradient with $3\sigma$ significance within a similar frequency interval, as shown in Fig.~\ref{fig:BH228A_RM}.
The detection suggests a high possibility of the existence of a helical \textbf{\textit{B}} field in the emission region of the core reflected by the frequency intervals (i.e. 5.0 $\sim$ 15.4 GHz in 2002 and 6.0 $\sim$ 12.2 GHz in 2020).
Although the helical field is believed common in the jet of AGN, only $\sim 50$ jets with firm transverse RM gradients are detected.
Most jets in AGN have their transverse structure unresolved and not enough polarized emission detected, which makes it difficult to obtain firm transverse RM gradients.
This needs a higher resolution and sensitivity array like space VLBI.
%For the observation in 2020, we do not detect transverse RM gradients.
%This means the jet is in the different condition from 2002.

When fitting RM, the RM corrected EVPA, which represents the intrinsic polarized direction and thus the projected structure of the \textbf{\textit{B}} field on the plane of the sky,  could also be obtained.
Fig.~\ref{fig:BH096_RM}, Fig.~\ref{fig:BH228A_RM} and Fig.~\ref{fig:BH228A_RM_V1} also provide the RM corrected EVPA for the two observations.
As can be seen, almost all the derived RM corrected EVPA distributions in the central region are along the jet direction except for that in 12.2-15.2 GHz in 2020, which shows a very large rotation of EVPA along the jet direction.
The corresponding RM distribution also shows a large change of magnitude with the further distance the smaller the values. 
This RM and intrinsic EVPA result in the observed polarization in the central region of 15.2 GHz in 2020 (as described in section 3.4.1 and shown in Fig.~\ref{fig:EVPA-AlongTheJet}).
The origin behind this change may lead to the evolution of the observed polarization at 15 GHz  detected by our observations and MOJAVE.

\section{discussion}
In this section, we discuss the evolution of the magnetic field for 1604+159 at pc scale.
\subsection{The magnetic field in the Jet}
MOJAVE monitored the source 1604+159 for 6 epochs from 2009 to 2013 at 15 GHz \citep{2018ApJS..234...12L} and studied the source with stacking \citep{2023MNRAS.520.6053P}.
Stacking has the advantage of improving the image sensitivity and providing a complete jet cross-section structure as much as possible by utilizing existing data.
Through distributions of EVPA and fractional polarization, stacking shows long-term persistent magnetic field configuration.
We revisited the stacking fits files derived by \citet{2023MNRAS.520.6053P} for 1604+159, clipped pixels which have polarization magnitude below 5$\sigma$ ($\sigma = 5\times10^{-5} \ \rm Jy \ beam^{-1}$ for this case), and obtained 4 transverse slices of fractional polarization roughly normal to the jet direction in the jet.
Fig.~\ref{fig:MOJAVE} plots the resultant slices.
Stacking results for this source have shown a stable linear polarization structure extending to $\sim4$ mas from the core for 1604+159 .
This polarization gradient basically orientates along the jet direction from the core to the jet, indicating a global toroidal \textbf{\textit{B}} field component existing in the jet.
In addition, the northern side of the jet has higher fractional polarization than that of the southern side, indicating a helical \textbf{\textit{B}} field rather than a toroidal \textbf{\textit{B}} field which could produce a symmetrical transverse fractional polarization profile.
If a helical \textbf{\textit{B}} field is in the jet, the side with the direction of \textbf{\textit{B}} field close to the LoS would have lower fractional polarization than the other side which has the direction of \textbf{\textit{B}} field close to the plane of the sky \citep{2011MNRAS.415.2081C}. 
%The higher fractional polarization stacked on the northern side of the jet than that on the southern side suggest that \textbf{\textit{B}} field go through out from the southern side of the jet and into the northern side assuming a global helical \textbf{\textit{B}} field in the jet.

Higher fractional polarization at one edge of the jet could also be explained by the interaction with ambient media \citep{2021Galax...9...58G}.
However, if a jet propagates through its ambient media, interaction with ambient media would produce a longitudinal \textbf{\textit{B}} field (i.e. orthogonal linear polarization) \citep{2014MNRAS.437.3405L}, which is not detected in the stable polarized gradient.
Thus, the helical \textbf{\textit{B}} field would be a more natural physical origin of the stable polarization structure.
But we could not rule out the interaction possibility.

According to the study of \citet{2013MNRAS.430.1504M}, given a helical \textbf{\textit{B}} field, the distribution of intrinsic EVPA depends on the pitch angle of the \textbf{\textit{B}} field and viewing angle.
As the jet of 1604+159 has low RM mediums (Fig.~\ref{fig:BH096_RM}, Fig.~\ref{fig:BH228A_RM} and Fig.~\ref{fig:BH228A_RM_V1}) surrounding the jet, the observed EVPA at frequencies above 4 GHz could be used to represent the intrinsic polarized direction.
The fact that the jet does not change its shape results in the expectation that the viewing angle would be unchanged between 2002 and 2020.
The pitch angle also appears unchanged.
Assuming a global helical magnetic field in the jet region, it is expected that a stable longitudinal EVPA distribution would be observed as the jet propagates basically along a line.
Comparing with the linear polarization results from 2009 to 2013 \citep{2023MNRAS.520.6053P} ,
we found that the polarization in the jet region changes with time, as can be seen in Fig.~\ref{fig:EVPA-AlongTheJet}.
Our results show that the polarization in the jet observed in  2002 and 2020 differs from the jet direction with a relatively large separation of $\sim \pm 40^{\circ}$ or even larger in some places.
We noted that the polarization of the jet observed in the two epochs is centered in bright, compact features, which could not be naturally explained as being associated with helical \textbf{\textit{B}} field \citep{2021Galax...9...58G}.
The large separation of EVPA suggests the helical \textbf{\textit{B}} field is severely distorted in the regions with compact features.
As discussed in section 3.3, shocks may exist in these regions with bright, compact features detected on the two epochs.
Shock compressing the \textbf{\textit{B}} field leads to the \textbf{\textit{B}} field along the shock front \citep{1980MNRAS.193..439L} .
With the relation of $90^{\circ}$ between the EVPA and the \textbf{\textit{B}} field in an optically thin area \citep{1970ranp.book.....P}, the polarization in the jet region in 2002 and 2020 indicates shocks are oblique, consistent with the study of \citet{2005AJ....130.1389L} who found the shocks could be oblique in quasars.
Our results support the point that shocks play an important role in the variability of jet emission, especially polarization variability \citep{2021Galax...9...58G}.

\subsection{The magnetic field in the central region}
While the central region at 15.4 GHz in 2002 remains EVPAs along the jet direction, it experiences a smooth change from perpendicular to parallel to the jet along the jet direction beginning at a place close to the jet base(Fig. \ref{fig:EVPA-AlongTheJet}).
This change has also been detected by MOJAVE at epoch 2009-Aug-19 and 2012-Jul-12 \footnote{\href{https://www.cv.nrao.edu/MOJAVE/sourcepages/1604+159.shtml}{https://www.cv.nrao.edu/MOJAVE/sourcepages/1604+159.shtml}}.

The Study of linear polarization variability \citep{2023MNRAS.523.3615Z} has shown that in the place close to the jet base, 1604+159 in $2009\sim2013$ has a large change of EVPA with time and the $\rm \sigma_{EVPA} \sim 50^{\circ}$.
This $\sigma$ is large in the population of quasars even in the blazars, suggesting the special physical condition of this source.
The significant difference in EVPA in the central region between 2002 and 2020 confirms the large $\sigma$.

Before exploring the origin of the significant evolution in EVPA, we investigate the potential magnetic field structure in the central region.
In section 4.1, we find evidence for the presence of a helical field in the jet region.
Here, we provide more observed evidence for the existence of a helical field in the core region.
First, the sign changed transverse RM gradients are detected in 2002 and 2020 (Fig.~\ref{fig:BH096_RM} and Fig.~\ref{fig:BH228A_RM}).
Second, the asymmetry of fractional polarization across the jet is reflected in Fig.~\ref{fig:MOJAVE}.
As pointed out by \citet{2013MNRAS.430.1504M}, the side when the fractional polarization (Fig.~\ref{fig:MOJAVE}) is lower has a larger magnitude of RM (Fig.~\ref{fig:BH096_RM} and Fig.~\ref{fig:BH228A_RM}), which we do observe in the southern side of the core.
Third, the RM corrected EVPA in the core has longitudinal distribution (Fig.~\ref{fig:BH096_RM}, Fig.~\ref{fig:BH228A_RM}, and Fig.~\ref{fig:BH228A_RM_V1}) except for the 12.2-15.2 GHz which will be discussed in the following content.
Compared with the results by \citet{2013MNRAS.430.1504M}, the longitudinal EVPA indicates a large intrinsic pitch angle ($\sim80^{\circ}$) in the jet rest frame assuming the polarized emission is dominated by optically thin regions.
Therefore, the potential magnetic field structure in the central region of 1604+159 could be a helical field with a large intrinsic pitch angle.

Under external screen assumption, the factors contributing to the evolution of observed EVPA at 15 GHz are RM and intrinsic EVPA (as indicated by equation \ref{RM}), both of which have a close relation with the $\textit{\textbf{B}}$ field.
 
In 2002, the 15 GHz related intrinsic EVPA shows distribution parallel to the jet direction (Fig.~\ref{fig:BH096_RM}).
However, in 2020, the intrinsic EVPA reflected by 12.2-15.2 GHz behaves more complex (Fig.~\ref{fig:BH228A_RM_V1}).
Here, we provide two possible explanations.
One is that, as discussed by \citet{2021Galax...9...58G}, the parallel relation between the EVPAs and projected $\textit{\textbf{B}}$ on the plane of the sky happens at an optical depth $\tau \backsimeq 6-7$ \citep{2018Galax...6....5W}, and the core observed in jet often locates at place having the optical depth $\sim$ 1 \citep{1979ApJ...232...34B}.
The observed polarized emission thus is from optically thin regions.
Therefore, if this is the case, the change in EVPA reflects a change of the inherent $\textit{\textbf{B}}$ field in the region of emission at 15 GHz.
This complex distribution indicates that the configuration of the helical magnetic field breaks in the region of emission at 15 GHz.

Another is that polarized emission from optically thick regions could not be negligible.
For the environments surrounding the jet, 15 GHz related RM in the core in 2020 is much larger than that in 2002 (Fig.~\ref{fig:BH096_RM} and Fig.~\ref{fig:BH228A_RM_V1}).
As the magnetic field strength measured in 2020 is $\sim$ 2 orders of magnitude smaller than that in 2002, the much larger RM in 2020 results from much higher electron density assuming the same integral region, which leads us to consider the contribution of polarized emission from optically thick regions.
A blending of emissions from both optically thin and thick regions results in the structure of the magnetic field being undetermined.
However, no matter which scenario is more appropriate, the change of intrinsic EVPA with time at the place close to the jet base reflects the change in the physical environment with time.
As the inner jet does not change its direction (Fig.~\ref{fig:Ridge Line} ), this large change (with time) of the $\textit{\textbf{B}}$ field structure may result from the rapid change of the environments surrounding the jet and/or central engine (e.g. the SMBH or/and the accretion disk) producing the $\textit{\textbf{B}}$ field.

Under the equipartition assumption, the core RM follows the relation $\rm|RM|$ $\propto \nu^{a}$, where a is the index in the electron density $n_{e}$ power-law change as a function of distance r from the black hole, $n_{e} \propto r^{-a}$ \citep{2007AJ....134..799J}.
The value of $a$ tells us the potential physics condition in the core.
For example, $a \sim 2$ is expected for a conically expanding jet having a boundary layer, which is the case of 1156+295, 2007+077, and 3C 273 \citep{2009MNRAS.393..429O,2019A&A...623A.111H}. 

We took the averaged RM and uncertainty in the central region as the core RM and its uncertainty for each frequency interval, listed in Table~\ref{tab:Core RM}.
Then we obtained the index $a = 2.7 \pm 0.5 $ from fitting, larger than 2, suggesting the core in 2020 has faster electron density fall-offs in the medium with distance from the jet base.
High index values are also observed in the core of 0954+658, 1418+546 \citep{2009MNRAS.393..429O}, 3C 279, OJ 287, CTA 102, 1749+096, 0235+164, and BL Lac \citep{2018ApJ...860..112P}.

The Large value of $a$ indicates some assumptions for a $\sim$ 2 break, including magnetic field structure and equipartition.
Although the value of 2.7 in 2020 does not significantly deviate from 2 if considering $2\sigma$ uncertainty, we do find other evidence that some assumptions break.
The magnetic field of the core reflected by the RM corrected EVPA distribution for 12.2-15.2 GHz in 2020 (Fig.~\ref{fig:BH228A_RM_V1}) has a more complex structure than that in 2002, a simple helical field.
The core shift index $k_{r}$ value of 0.32 suggests the extent of deviation from equipartition is more significant than in 2002.

We have observed significant behavior differences in linear polarization at 15 GHz and in the Faraday Rotation medium in the core in the two epochs.
These variations indicate that the \textit{B} field wound up by the central engine in the core correlates with the medium wrapping up it.
Linear polarization at 15 GHz collected spanning $\sim 18$ years from 2002 to 2020 suggests the core of 1604+159 experiences a complex evolution of physical condition including the \textbf{\textit{B}} field and the medium near the jet base.

Other blazars are showing significant evolution of \textbf{\textit{B}} field and RM.
\citet{2009MNRAS.393..429O} found no good linear $\lambda^{2}$ fits for the source 1749+096 observed in 2006 July, while observation two weeks earlier detected core RM \citep{2012AJ....144..105H}.
One possible explanation for this variation is a possible flare in the source affecting the polarization results.
\citet{2009MNRAS.400....2M} found a surprising reversal of Faraday Rotation gradients in the jet of the BL Lac object B1803+784.
One of the most likely origins they explained is a 'nested-helix' magnetic field structure, resulting from magnetic-power-type scenarios \citep{1996MNRAS.279..389L}.
The origins of RM change with time are different for these sources.
Whether the differences are common in AGN needs studies of the evolution of RM with a large number of sources.

\section{Conclusions}

We have analyzed the total intensity, spectral index, linear polarization, and RM distributions at pc scale for the quasar 1604+159.
The source was observed at 5.0, 8.4, and 15.4 GHz  in 2002, and 4.6, 5.1, 6.0, 7.8, 12.2, 15.2, and 43.9 GHz  in 2020 with the VLBA.
Combining the MOJAVE polarization results at 15 GHz from 2009 to 2013, we studied the evolution of the magnetic field spanning $\sim 18$ years.
Based on the linear polarization distribution in 2002, we divided the source structure into the central region and the jet region.
The conclusions are summarized as follows:

\begin{enumerate}
    \item We detected a core-jet structure.
The jet extends to the distance of $\sim 25$ mas from the core at\textbf{ }a direction of $\sim 66^{\circ}$ north by east.
The shape of the jet derived from 15 GHz data varies slightly with time and could be described by a straight line.
\item The core shift index $k_{r}$ in 2002 indicates the core is close to equipartition, while in 2020 the core shows a large deviation.
The measured magnetic field strength in 2020 is two orders of magnitude lower than that in 2002.
    \item We find the polarized emission in the jet region varies obviously with time.
The polarization has a large direction separation of $\sim \pm40^{\circ}$ from the jet direction in the two epochs.
These polarized emissions are centered in regions with bright, compact jet features, which have spectral index values along the jet ridge larger than the mean spectral index of quasars in the jet.
The flatter spectral index values indicate the possible existence of shocks in these bright, compact jet features, contributing to the variation of polarization in the jet with time.
The oblique EVPAs indicate the shocks are oblique.
    \item In the central region, the spectral index shows flat spectra and the fractional polarization is low compared with that in the jet region.
The polarized emission orientates in general towards the jet direction.
At 15 GHz, in the place close to the jet base, the EVPA changes significantly with time from perpendicular to parallel to the jet direction.
We find in 2002 the EVPA at 15 GHz in this place is basically along the jet and the rotation of that at the three frequencies could be well fitted to a linear model (external medium screen).
In 2020, the EVPA at this place becomes perpendicular, and the rotation of that behaves more complex.
\item We detected transverse RM gradients for the two observations in the core, suggesting the existence of a helical field. 
The 15 GHz related RM in the core in 2020 shows a large change with distance.
The RM corrected EVPA distribution in 12.2-15.2 GHz in 2020 reflects a complex magnetic field structure in the core of 15 GHz.
The core $\rm |RM|$ increases with frequency following a power law with index $a = 2.7 \pm 0.5 $, showing faster electron density fall-offs in the medium with distance from the jet base than a sheath surrounding a conically expanding jet in which the power law index is $\sim 2$. 
\end{enumerate}

\section*{Acknowledgements}
We thank the referee for the comments improving the manuscript.
This work was supported by the National Key R$\&$D Programme of China (2018YFA0404602).
L.C. appreciates the support from the National Natural Science Foundation of China (grant 12173066).
The VLBA data of the C band and Ku band of BH228A were calibrated using NRAO's VLBA data calibration pipeline in AIPS.
This research has made use of data from the MOJAVE database that is maintained by the MOJAVE team \citep{2018ApJS..234...12L}.


\vspace{5mm}
\facility{VLBA}


\software{astropy \citep{2013A&A...558A..33A,2018AJ....156..123A,2022ApJ...935..167A}
    }



%\section{Using Chinese, Japanese, and Korean characters}

%Authors have the option to include names in Chinese, Japanese, or Korean (CJK) 
%characters in addition to the English name. The names will be displayed 
%in parentheses after the English name. The way to do this in AASTeX is to 
%use the CJK package available at \url{https://ctan.org/pkg/cjk?lang=en}.
%Further details on how to implement this and solutions for common problems,
%please go to \url{https://journals.aas.org/nonroman/}.


\bibliography{sample631}{}
\bibliographystyle{aasjournal}



\end{document}